\documentclass[apj,numberedappendix]{emulateapj}
\slugcomment{Accepted version; December 2017}

\pdfoutput=1
\usepackage[backref,breaklinks,colorlinks,citecolor=blue]{hyperref}
\usepackage[all]{hypcap} 

\usepackage{amsmath,tabularx, rotating, booktabs}
\newcommand\ddfrac[2]{\frac{\displaystyle #1}{\displaystyle #2}}

\usepackage{lipsum}
\graphicspath{{./smuvs_clustering_paper_plots/}}

\shorttitle{The Galaxy-Halo Connection for $1.5\lesssim\lowercase{z}\lesssim5$ in SMUVS}
\shortauthors{Cowley, W.I. et al.}

\usepackage{layouts}

\begin{document}

\title{The Galaxy-Halo Connection for $1.5\lesssim\lowercase{z}\lesssim5$ as revealed by the \emph{Spitzer} Matching survey of the UltraVISTA ultra-deep Stripes}

\author{
William~{I.}~Cowley\altaffilmark{1}, 
Karina~I.~Caputi\altaffilmark{1},
Smaran~Deshmukh\altaffilmark{1},
Matthew~{L.N.}~Ashby\altaffilmark{2},
Giovanni~{G.}~Fazio\altaffilmark{2},
Olivier~Le~F\`evre\altaffilmark{3},
Johan~{P.U.}~Fynbo\altaffilmark{4},
Olivier~Ilbert\altaffilmark{3},
Henry~{J.}~McCracken\altaffilmark{5},
Bo~Milvang-Jensen\altaffilmark{4}, and
Rachel~{S.}~Somerville\altaffilmark{6}
}

\altaffiltext{1}{Kapteyn Astronomical Institute, University of Groningen, P.O. Box 800, 9700 AV, Groningen, The Netherlands; \href{mailto:cowley@astro.rug.nl}{cowley@astro.rug.nl}}
\altaffiltext{2}{Harvard-Smithsonian Center for Astrophysics, 60 Garden Street, Cambridge, MA 02138, USA}
\altaffiltext{3}{Aix Marseille Universit\'e, CNRS, LAM (Laboratoire d'Astrophysique de Marseille) UMR 7326, F-13388, Marseille, France}
\altaffiltext{4}{Dark Cosmology Centre, Niels Bohr Institute, University of Copenhagen, Juliane Maries Vej 30, DK-2100 Copenhagen, Denmark}
\altaffiltext{5}{Institut d'Astrophysique de Paris, CNS \& UPMC, UMR 7095, 98 bis Boulevard Arago, F-75014, Paris, France}
\altaffiltext{6}{Department of Physics and Astronomy, Rutgers University, The State University of New Jersey, 136 Frelinghuysen Road, Piscataway, NJ 08854 USA}

\begin{abstract}
The \emph{Spitzer} Matching Survey of the UltraVISTA ultra-deep Stripes (SMUVS) provides unparalleled depth at $3.6$ and $4.5$~$\mu$m over $\sim0.66$~deg$^2$ of the COSMOS field, allowing precise photometric determinations of redshift and stellar mass. From this unique dataset we can connect galaxy samples, selected by stellar mass, to their host dark matter halos for $1.5<z<5.0$, filling in a large hitherto unexplored region of the parameter space. To interpret the observed galaxy clustering we utilize a phenomenological halo model, combined with a novel method to account for uncertainties arising from the use of photometric redshifts. We find that the satellite fraction decreases with increasing redshift and that the clustering amplitude (e.g., comoving correlation length / large-scale bias) displays monotonic trends with redshift and stellar mass. Applying $\Lambda$CDM halo mass accretion histories and cumulative abundance arguments for the evolution of stellar mass content we propose pathways for the coevolution of dark matter and stellar mass assembly. Additionally, we are able to estimate that the halo mass at which the ratio of stellar to halo mass is maximized is $10^{12.5_{-0.08}^{+0.10}}$~M$_{\odot}$ at $z\sim2.5$. This peak halo mass is here inferred for the first time from stellar mass-selected clustering measurements at $z\gtrsim2$, and implies mild evolution of this quantity for $z\lesssim3$, consistent with constraints from abundance-matching techniques.
\end{abstract}


\section{Introduction}
\label{sec:intro}
A central \emph{ansatz} of the $\Lambda$ cold dark matter ($\Lambda$CDM) cosmological paradigm is that galaxies form from baryonic condensations within the potential well of a dark matter halo \citep[e.g.,][]{WhiteRees:1978}. However, understanding the relationship between a dark matter halo and the galaxies it hosts is far from a trivial exercise. In particular, the issue this paper explores is how the stellar mass of a galaxy is related to the mass of its host dark matter halo, and how this relationship evolves with cosmic time.  

A conceptually straightforward way in which to investigate the relation between galaxies and their host halos is a direct simulation of galaxy formation within a cosmological context. The current state of the art of such efforts comprises an $N$-body computation of the evolution of dark matter combined with either a hydrodynamical \citep[e.g.,][]{Vogelsberger:2014,Schaye:2015} or semi-analytical \citep[e.g.,][]{Somerville:2012,Henriques:2015,Lacey:2016} treatment of the baryonic processes involved. Both schemes have enjoyed considerable and groundbreaking successes in reproducing multiple observational datasets, thereby furthering our understanding of galaxy formation and evolution.  However, they are both hampered by a lack of our knowledge regarding some of the complex physical processes involved, such as star and black hole formation, and associated feedback processes. This is largely related to the extreme computational challenge posed by attempting to resolve these processes in simulations of cosmologically significant volumes and time periods. As a result, some uncertainties regarding how accurately the physical processes involved in producing galaxies within dark matter halos are accounted for in these models remain.      

Direct observational probes of the dark matter halos of galaxies include galaxy-galaxy lensing \citep[e.g.,][]{Brainerd:1996,Hoekstra:2004,Leauthaud:2010,Hudson:2015} and the kinematics of satellite galaxies \citep[e.g.,][]{Zaritsky:1993,BrainerdSpecian:2003,vdBosch:2004,Norberg:2008,More:2011}. Galaxy-galaxy lensing uses distortions of the shapes and orientations of background galaxies caused by intervening mass along the line of sight and thus can be used to infer the foreground mass distribution. Satellite kinematics uses satellite galaxies as test particles in order to trace out the dark matter velocity field, and thus potential well, of the dark matter halo.

However, both of these techniques are mostly limited to low redshifts ($z<1$), due to the difficulty of resolving individual galaxies at earlier epochs, and they rely on ensemble stacking of galaxies in order to extract a signal.

A simpler, though less direct, approach is to compare the observed abundance and clustering properties of galaxy samples with predictions from a phenomenological halo model \citep[e.g.,][]{NeymanScott:1952,BerlindWeinberg:2002,CooraySheth:2002}. This is a purely statistical description of how galaxies occupy halos and thus forgoes an understanding of the physical processes involved. However, this technique has been shown to provide a good description of the observed clustering of galaxies and can in principle be applied over very broad redshift ranges. 

A drawback of this approach is that it relies on accurate \emph{a priori} knowledge of the matter power spectrum, halo mass function, halo density profile, and the bias with which halos trace the underlying matter distribution. The relations used in practice are calibrated against numerical simulations, though there remains some doubt over the applicable regime for these calibrations. It also assumes that the bias with which halos trace the underlying matter distribution depends only on halo mass, ignoring potential effects such as `halo assembly bias' \citep[e.g.,][]{Zenter:2014}.

In recent years the halo occupation modeling technique has been utilized in many studies investigating the galaxy-halo connection \citep[e.g.,][]{Zheng:2004,Zheng:2007,Zehavi:2011,Bethermin:2014,Skibba:2015}, and some works have combined this technique with constraints from galaxy-galaxy lensing \citep[e.g.,][]{Leauthaud:2010,Coupon:2015}. However, interpreting the results from such studies in terms of physical galaxy properties is often complicated by the selection of the galaxies used. Many samples are selected by luminosity, resulting in uncertainty regarding the conversion to stellar mass. 

Another difficulty is probing a sufficient range of halo masses such that the galaxy-halo relation can be meaningfully characterized. Increasing the depth of a survey, allowing lower mass galaxies to be investigated, often limits the survey area such that the number of massive galaxies is not sufficient for a robust clustering analysis. This trade-off has proven difficult to overcome for $z\gtrsim1$. 

Galaxy clustering at higher redshifts (up to $z\sim7$) has been investigated through use of the Lyman break `dropout' selection method \citep[e.g.,][]{Giavalisco:1998,Harikane:2016,Hatfield:2017,Ishikawa:2017} and Lyman~$\alpha$ emission \citep[e.g.,][]{Ouchi:2017}. However, as in these works galaxies are selected by broadband color and nebular emission respectively, the connection to properties such as stellar mass is even more uncertain.                 

Recently, \cite{McCracken:2015} provided a clustering analysis of galaxies selected by stellar mass up to $z\sim2$ based on UltraVISTA DR1 near-infrared and ancillary COSMOS broadband data \citep{McCracken:2012}. These authors were able to directly characterize the stellar-to-halo mass relation (hereafter SHMR) up to $z\sim2$.       

Here, we use the unique \emph{Spitzer} Matching survey of the UltraVISTA ultra-deep Stripes (SMUVS) galaxy catalog (Ashby et al. \citeyear{Ashby:2018}; Deshmukh et al. \citeyear{Deshmukh:2018}) to extend our understanding of the relationship between a galaxy's stellar mass and its host dark matter halo to higher redshifts than were previously possible. We do so through comparing the observed clustering and abundances of stellar mass-selected samples of galaxies to predictions from a phenomenological halo occupation model.

This paper is structured as follows: in Section~\ref{sec:smuvs_survey} we describe the galaxy catalog used throughout this work and the construction of our galaxy samples. In Section~\ref{sec:methods} we describe the phenomenological halo model used to interpret our observations, and introduce a novel method to account for the use of photometric redshifts (this is discussed in more detail in Appendix~\ref{sec:delta_z_discuss}). We present our results in Section~\ref{sec:results} (our main results are tabulated in \autoref{table:main_results}, Appendix~\ref{sec:main_results_table}). A brief discussion of some of the modeling assumptions made in our analyses is given in Section~\ref{sec:discussion}. We conclude in Section~\ref{sec:conclusion}. 

Throughout we assume a flat $\Lambda$CDM cosmology with $\Omega_{\rm m}=0.3$, $\Omega_{\Lambda}=0.7$, $\Omega_{\rm b}=0.045$, $h=0.7$, $\sigma_{8}=0.8$, and $n_{\rm s}=0.95$. All magnitudes are quoted in the Absolute Bolometric (AB) system \citep{Oke:1974}.    
\section{The SMUVS survey}
\label{sec:smuvs_survey} 
\subsection{Survey overview} 
The SMUVS program (PI: K. Caputi; Ashby et al. \citeyear{Ashby:2018}) has collected ultra-deep \emph{Spitzer} $3.6$ and $4.5$~$\mu$m data over the region of the COSMOS\footnote{\url{http://cosmos.astro.caltech.edu}} field \citep{Scoville:2007} overlapping with three of the UltraVISTA ultra-deep stripes \citep{McCracken:2012} with deep optical coverage from the Subaru Telescope \citep{Taniguchi:2007}. The UltraVISTA data considered here correspond to the third data release\footnote{\url{http://www.eso.org/sci/observing/phase3/data_releases/uvista_dr3.pdf}}, which reaches an average depth of $K_{\rm s}=24.9\pm0.1$ and $H=25.1\pm0.1$ ($2$~arcsec diameter, $5\sigma$).  This paper forms part of a series of scientific studies that make use of SMUVS data, such as the search for strong H$\alpha$ emitters at $z=4-5$ \citep{Caputi:2017} and the study of galaxy structural properties for $z\lesssim5$ \citep{Hill:2017}.   

A thorough description of the SMUVS multiwavelength source catalog construction and spectral energy distribution (SED) fitting is given in \cite{Deshmukh:2018}, however, we summarize the main details here.

Sources are extracted from the UltraVISTA $HK_{\rm s}$ average stack mosaics using {\sc SExtractor} \citep{BertinArnouts:1996}. The positions of these sources were then used as priors to perform iterative point-spread function (PSF) fitting photometric measurements on the SMUVS $3.6$ and $4.5$~$\mu$m mosaics, using the {\sc daophot} package \citep{Stetson:1987}.

For all of these sources, $2$~arcsec diameter circular photometry on $26$ broad, intermediate, and narrow bands $U$ to $K_{\rm s}$ is measured \citep{Deshmukh:2018}. After cleaning for galactic stars using a $B$-$J$-$[3.6]$ color selection \citep[e.g.,][]{Caputi:2011}, and masking regions of contaminated light around the brightest sources, the final catalog contains $\sim2.9\times10^5$ UltraVISTA sources with a detection in at least one IRAC band over an area of $\sim0.66$ square degrees.

The SED fitting is performed with all $28$ bands ($26$ $U$ through $K_{\rm s}$ as well as \emph{Spitzer} $3.6$ and $4.5$~$\mu$m) using the $\chi^2$ minimization code {\sc LePhare}\footnote{\url{http://www.cfht.hawaii.edu/~arnouts/LEPHARE/lephare.html}} \citep{Arnouts:1999,Ilbert:2006}. We assume \cite{BruzualCharlot:2003} templates corresponding to a simple stellar population formed with a \cite{Chabrier:2003} stellar initial mass function and either solar or sub-solar ($1$~Z$_{\odot}$ or $0.2$~Z$_{\odot}$) metallicity, and allow for the addition of nebular emission lines. Additionally, we assume exponentially declining star formation histories. 

Photometric redshifts and stellar mass estimates are obtained for $>99$~percent of our sources. Using ancillary spectroscopic data in COSMOS to assess the quality of the obtained photometric redshifts, we found that the standard deviation, $\sigma_{z}$, of $|z_{\rm phot}-z_{\rm spec}| / (1 + z_{\rm spec})$, based on $\sim1.4\times10^4$ galaxies with reliable spectroscopic redshifts in the COSMOS field (see Table~1 in Ilbert et al., \citeyear{Ilbert:2013} and references therein) is $0.026$ \citep{Deshmukh:2018}. This statistic is computed excluding outliers, defined as objects for which $|z_{\rm phot}-z_{\rm spec}| / (1 + z_{\rm spec})>0.15$, which comprise $\sim5.5$~percent of the spectroscopic catalog. These results compare favorably with other photometric surveys in the literature \citep[e.g.][]{Ilbert:2013,Laigle:2016} and highlight the high accuracy of our derived photometric redshifts.      

Throughout we use the best-fit redshifts and stellar masses computed by {\sc LePhare}.
\subsection{Sample selection}
\begin{figure}
\centering
\includegraphics[width = 8.6189cm]{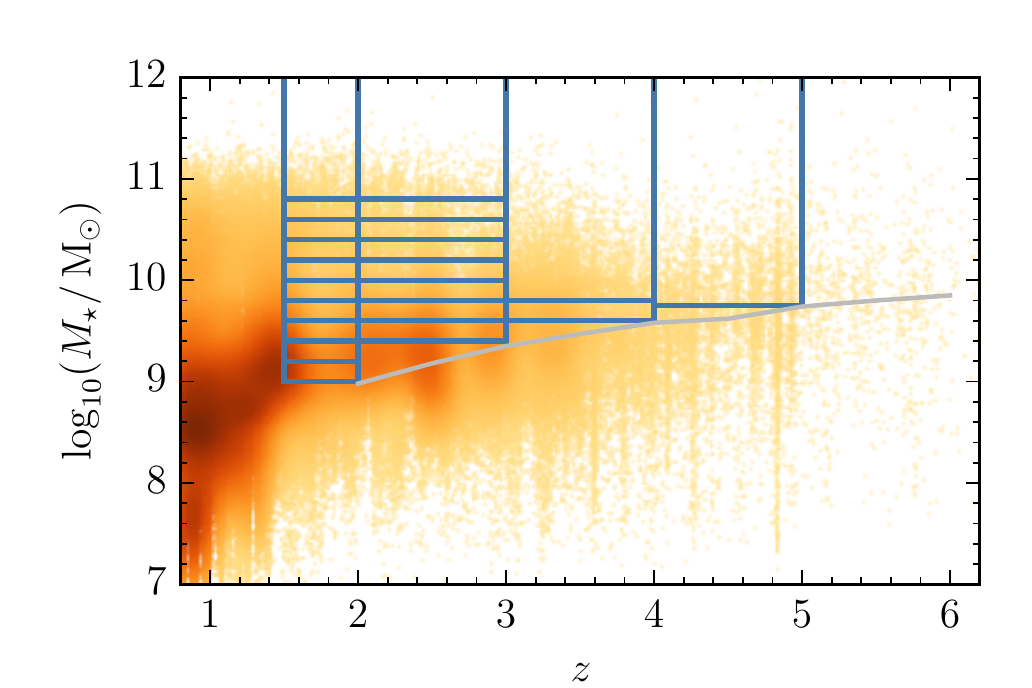}
\caption{The stellar mass -- redshift plane for SMUVS galaxies. The dark blue lines indicate our stellar mass-selected volume-limited samples. The light gray line indicates the $80$~percent stellar mass completeness limit.  The color scale indicates the density of objects at that point on the plane (darker color indicates a higher density).}
\label{fig:mstar_v_z}
\end{figure} 
\capstartfalse
\begin{deluxetable*}{rrrrrrrrr}
\tablewidth{0.8\linewidth}
\tablecaption{Characteristics of each galaxy sample}
\tablehead{
& \multicolumn{2}{c}{$1.5<z<2.0$} & \multicolumn{2}{c}{$2.0<z<3.0$} & \multicolumn{2}{c}{$3.0<z<4.0$} & \multicolumn{2}{c}{$4.0<z<5.0$} \\
\colhead{Threshold$^{\mathrm{(a)}}$} & \colhead{$N_{\mathrm{gal}}$} & \colhead{$M_{\star\mathrm{,}50}^{\mathrm{(a)}}$} & \colhead{$N_{\mathrm{gal}}$} & \colhead{$M_{\star\mathrm{,}50}^{\mathrm{(a)}}$} & \colhead{$N_{\mathrm{gal}}$} & \colhead{$M_{\star\mathrm{,}50}^{\mathrm{(a)}}$} & \colhead{$N_{\mathrm{gal}}$}} & \colhead{$M_{\star\mathrm{,}50}^{\mathrm{(a)}}$}
\startdata
$ 9.00$&$19\,637$&$ 9.49$&--&--&--&--&--&--\\
$ 9.20$&$15\,309$&$ 9.64$&--&--&--&--&--&--\\
$ 9.40$&$11\,232$&$ 9.86$&$18\,362$&$ 9.73$&--&--&--&--\\
$ 9.60$&$ 8\,186$&$10.09$&$12\,201$&$ 9.92$&$ 6\,187$&$ 9.89$&--&--\\
$ 9.75$&--&--&--&--&--&--&$ 1\,768$&$10.02$\\
$ 9.80$&$ 6\,129$&$10.27$&$ 7\,851$&$10.14$&$ 3\,894$&$10.04$&--&--\\
$10.00$&$ 4\,634$&$10.42$&$ 5\,194$&$10.35$&--&--&--&--\\
$10.20$&$ 3\,458$&$10.54$&$ 3\,516$&$10.52$&--&--&--&--\\
$10.40$&$ 2\,398$&$10.66$&$ 2\,357$&$10.66$&--&--&--&--\\
$10.60$&$ 1\,469$&$10.79$&$ 1\,446$&$10.81$&--&--&--&--\\
$10.80$&$  702$&$10.95$&$  740$&$10.93$&--&--&--&--
\enddata
\label{table:samp_characteristics}
\tablecomments{$^{\mathrm{(a)}}$ in $\log_{10}(M_{\star}/\mathrm{M}_{\odot})$. For each stellar mass threshold and redshift bin we report the number of galaxies and the median stellar mass.}
\end{deluxetable*}
\capstarttrue
We construct volume-limited stellar mass-selected samples of galaxies for this analysis. The stellar mass--redshift plane for the SMUVS catalog is shown in \autoref{fig:mstar_v_z}. The $80$~percent stellar mass completeness limit (based on $4.5$~$\mu$m photometry) is calculated following the method described in Chang et al. (\citeyear{Chang:2013}, see also Tomczak et al. \citeyear{Tomczak:2014}) and is shown as the gray line. All of our samples are above this completeness limit. The number of objects in each sample, as well as the median stellar mass, is summarized in \autoref{table:samp_characteristics}.

We restrict our analysis to $z>1.5$ as the COSMOS field has a well-documented overabundance of rich structures \citep{McCracken:2007, Meneux:2009, McCracken:2015} at $z\sim1-1.5$ that can nullify the halo modeling analysis performed in this work, and we wish to focus on the high-redshift nature of our catalog. Halo modeling analyses have been performed for $z\lesssim1$ in many previous studies \citep[e.g.][]{Leauthaud:2011,Coupon:2012,McCracken:2015} which cover a larger volume at these redshifts than SMUVS.          
\section{Methods}
\label{sec:methods}
\subsection{The angular two-point correlation function}
The angular autocorrelation function, $w(\theta)$, describes the excess probability, compared to a random (Poisson) distribution, of finding a pair of galaxies at some angular separation $\theta > 0$. It is defined such that
\begin{equation}
\delta^{2}P_{12} = \bar{\eta}^{2}\,[1 + w(\theta)]\,\delta\Omega_{1}\,\delta\Omega_{2}\rm,
\end{equation}
where $\bar{\eta}$ is the mean surface density of the population per unit solid angle, $\theta$ is the angular separation, and $\delta\Omega_{\rm i}$ is a solid angle element.
Here, the angular correlation function is computed according to the standard \cite{LandySzalay:1993} estimator
\begin{equation}
w(\theta) = \frac{DD(\theta) - 2DR(\theta) + RR(\theta)}{RR(\theta)}\rm,
\label{eq:wtheta}
\end{equation}
where $DD$, $DR$ and $RR$ represent the number of data-data, data-random, and random-random pairs in a bin of angular separation respectively, and the random catalog is constructed to have the same angular selection as the data. We use a random catalog with $\sim5\times10^5$ objects. The errors on the two-point angular correlation function are estimated from the data using a jackknife approach \citep{Norberg:2009}. We divide the SMUVS footprint into $60$ approximately equal area jackknife regions, removing one region at a time we compute the covariance matrix as 
\begin{equation}
C_{i,j} = \frac{N-1}{N}\sum_{l=1}^{N}(w_{i}^{l}-\bar{w_{i}})\times(w_{j}^{l}-\bar{w_{j}})\rm,
\label{eq:covar_matrix}
\end{equation}
where $N$ is the total number of jackknife regions, $\bar{w}$ is the mean correlation function ($\sum_{l}w^{l}/N$) and $w^{l}$ is the estimate of $w$ with the $l^{\rm th}$ region removed.  We compute $w(\theta)$ for each stellar mass threshold and redshift bin described in \autoref{table:samp_characteristics}, for $12$ evenly spaced logarithmic bins of angular separation in the range $-3<\log_{10}(\theta/\mathrm{deg})<-0.6$.
\subsection{The halo occupation model}
We use an established phenomenological halo occupation model \citep{Zheng:2007} to connect the observed galaxy angular correlation functions and abundances to host dark matter halos. In the halo model, the expected mean number of galaxies in a dark matter halo, $N(M_{\rm h})$, the halo occupation distribution (HOD), is a sum of contributions from central galaxies, $N_{\rm c}$, and satellites, $N_{\rm s}$, such that
\begin{equation}
N(M_{\rm h}) = N_{\rm c}(M_{\rm h})\times[1 + N_{\rm s}(M_{\rm h})]\rm.
\end{equation}
The contribution from central galaxies (i.e., those at the center of the halo potential well) is modeled as a step function with a smooth transition: 
\begin{equation}
N_{\rm c}(M_{\rm h})=\frac{1}{2}\left[1 + \mathrm{erf}\left(\frac{\log M_{\rm h} - \log M_{\rm h,min}}{\sigma_{\log M_{\rm h}}} \right)\right]\rm ,
\label{eq:hod_ncen}
\end{equation}
where $M_{\rm h,min}$ is the mass at which $50$~percent of halos host a single galaxy and $\sigma_{\log M_{\rm h}}$ is the width of the central galaxy mean occupation. The contribution from satellite galaxies is modeled as a power law with a cutoff at low halo masses:
\begin{equation}
N_{\rm s}(M_{\rm h}) = \left(\frac{M_{\rm h} - M_{\mathrm{h,}0}}{M_{\mathrm{h,}1}}\right)^{\alpha_{\rm sat}}\rm,
\end{equation}
where $M_{\mathrm{h,}0}$ is the cutoff mass scale, $M_{\mathrm{h},1}$ characterizes the amplitude, and $\alpha_{\rm sat}$ describes the asymptotic slope at high halo mass.
 
Thus, the HOD is described by five parameters: $M_{\rm h,min}$, $M_{\mathrm{h,}1}$, $M_{\mathrm{h,}0}$, $\alpha_{\rm sat}$ and $\sigma_{\log M_{\rm h}}$. 

A spatial correlation function, $\xi(r)$, where $r$ is the comoving spatial separation, is then constructed from the HOD\footnote{This is later projected for comparison with the measured angular correlation functions as described in Section~\ref{sec:project_photz}.}. This step assumes a number of relations, generally calibrated against numerical simulations, which are described in the following two paragraphs.

A Navarro-Frenk-White \citep[NFW; ][]{NFW:1997} halo density profile, with the concentration relation of \cite{Bullock:2001} is assumed. The halo mass function used is the parametrization of \cite{Tinker:2008}, with the high-redshift correction of \cite{Behroozi:2013b}, as well as the large-scale dark matter halo bias parametrization of \cite{Tinker:2010}. These Tinker et al. relations adopt the definition of a halo as a spherical overdensity of $200$ relative to the mean cosmic density at the epoch of interest \citep[e.g.][]{LaceyCole:1994, Tinker:2008}.

The matter power spectrum is computed according to \cite{EisensteinHu:1995} with the nonlinear correction of \cite{Smith:2003} applied. Additionally, we implement the two halo exclusion model of \cite{Tinker:2005}, which improves on that presented in \cite{Zheng:2004}. A thorough description of the construction of a similar halo model, upon which the one used in this work is based, is given in Appendix~A of \cite{Coupon:2012}.

Given an HOD we can compute the following derived quantities which we will discuss later: the satellite fraction, $f_{\rm sat}$,
\begin{eqnarray}
f_{\rm sat}(z) &= & 1 - f_{\rm cen}(z)\nonumber \\
 &=& 1 - \int N_{\rm c}(M_{\rm h},z)\,n(M_{\rm h},z)\,\mathrm{d}M_{\rm h}\,/\,n_{\rm gal}(z)\rm,
\label{eq:fsat}
\end{eqnarray}
where $n(M_{\rm h},z)$ is the halo mass function and $n_{\rm gal}$ is the galaxy number density,
\begin{equation}
n_{\rm gal}(z) = \int N(M_{\rm h})\,n(M_{\rm h}, z)\,\mathrm{d}M_{\rm h}\rm,
\label{eq:ngal}
\end{equation}
and the effective large-scale galaxy bias, $b_{\rm gal}$,
\begin{equation}
b_{\rm gal}(z) = \int b_{\rm h}(M_{\rm h}, z)\,N(M_{\rm h})\,n(M_{\rm h},z)\,\mathrm{d}M_{\rm h}\,/\,n_{\rm gal}(z)\rm,
\label{eq:bias}
\end{equation}
where $b_{\rm h}(M_{\rm h}, z)$ is the large-scale halo bias parametrization of \cite{Tinker:2010}.
\subsection{Projection and the effect of photometric redshift errors}
\label{sec:project_photz}
In this work we measure angular correlation functions of galaxies in bins of redshift. This is a necessary consequence of our photometric redshifts, which are not accurate enough to measure the galaxy distribution in three dimensions. Thus a spatial correlation function, $\xi(r)$, computed from an HOD, needs to be projected along the line of sight into two dimensions for comparison with our data. For this we use the \cite{Limber:1953} equation,
\begin{equation}
w(\theta) = \ddfrac{\int \left(n_{\rm gal}(z)\,\frac{\mathrm{d}V}{\mathrm{d}z}\,W(z)\right)^{2}\frac{\mathrm{d}z}{\mathrm{d}\chi}\,\mathrm{d}z\int\xi(r,z)\,\mathrm{d}u}{\left(\int n_{\rm gal}(z)\,\frac{\mathrm{d}V}{\mathrm{d}z}\,W(z)\,\mathrm{d}z\right)^{2}}\rm,
\label{eq:limber}
\end{equation}
where $n_{\rm gal}$ is the number density of galaxies predicted by the HOD, $\mathrm{d}V/\mathrm{d}z$ is the comoving volume element, $\mathrm{d}z/\mathrm{d}\chi=H_{0}E(z)/c$ where $E(z)=[\Omega_{\mathrm{m}}(1+z)^{3} + \Omega_{\Lambda}]^{1/2}$, and $\chi$ corresponds to the comoving radial distance to redshift $z$.  The comoving line-of-sight separation, $u$, is defined by $r=[u^{2} + \chi^{2}\varpi^{2}]^{1/2}$ where $\varpi^{2}/2 = [1 - \cos(\theta)]$. The $W(z)$ term in \autoref{eq:limber} relates to the redshift window that is being probed by the survey. If the redshifts were known precisely, then (ignoring further complications such as redshift space distortions) this would be a top-hat function equal to unity between the limits of the redshift range probed and zero elsewhere. However, with photometric redshifts, this is not the case, and the top-hat window should be convolved with an error kernel that is generally unknown \emph{a priori}.

In order to mitigate this, here we assume a Gaussian error kernel and approximate a $(1+z)$ evolution in the error kernel dispersion, $\Delta_{z}$, such that
\begin{equation}
W(z) = \frac{1}{2}\left[\mathrm{erf}\left(\frac{z-z_{\rm lo}}{\Delta_{z}\,(1+z_{\rm lo})}\right) - \mathrm{erf}\left(\frac{z-z_{\rm hi}}{\Delta_{z}\,(1+z_{\rm hi})}\right) \right]\rm.
\label{eq:sigma_z_window}
\end{equation}
Here $z_{\rm lo}$ and $z_{\rm hi}$ represent the lower and upper redshift limits of the photometric redshift bin respectively. Whilst the integral in \autoref{eq:limber} is, in principle, over all redshift, it only has significant contributions from redshifts where $W(z)$ is appreciably nonzero. We leave $\Delta_{z}$ as a free parameter in our fitting procedure, which is described in Section~\ref{sec:fitting}. This decision, and our modeling of the photometric redshift dispersion, is discussed in more detail in Appendix~\ref{sec:delta_z_discuss}.

Once projected according to \autoref{eq:limber}, we account for the integral constraint \citep[e.g.,][]{GrothPeebles:1977}. This is the correction required as the measured angular correlation function will integrate to zero over the whole field, by construction of the estimator for $w(\theta)$ used. This results in the measured angular correlation function being underestimated by an average amount
\begin{equation}
\sigma_{\rm IC}^{2} = \frac{1}{\Omega^{2}}\int\int w_{\rm true}(\theta)\,\mathrm{d}\Omega_{1}\mathrm{d}\Omega_{2}\rm,
\label{eq:integral_constraint}
\end{equation}
where $w_{\rm true}$ is the true angular correlation function and the angular integrations are performed over a field of area $\Omega$.

Here we evaluate \autoref{eq:integral_constraint} according to the numerical method proposed by \cite{RocheEales:1999},
\begin{equation}
\sigma_{\rm IC}^{2} = \frac{\sum w_{\rm true}(\theta)\,RR(\theta)}{\sum RR(\theta)}\rm.
\end{equation}
We take $w_{\rm true}$ to be the angular correlation function predicted by our HOD model, and subtract $\sigma_{\rm IC}^{2}$ from it before comparing it to our observed correlation functions. We do this for each evaluation of $w(\theta)$ in our fitting procedure, which is described below.  These corrections are typically $<10^{-2}$.
\subsection{Fitting}
\label{sec:fitting}
We derive the best-fitting halo model corresponding to our clustering and abundance measurements using the {\sc python} affine-invariant implementation for Markov chain Monte Carlo (MCMC), {\sc emcee}\footnote{\url{http://dan.iel.fm/emcee/}} \citep{Foreman-Mackey:2013}. Although our model in principle has six adjustable parameters, after some initial tests we determined that, despite a good signal being measured in our correlation functions, our data were insufficient to fully constrain all of them. This was based on consideration of the Bayesian information criterion (BIC; Schwarz \citeyear{Schwarz:1978}; see e.g., \citealt{Liddle:2007} for a discussion of this criterion). It is an approximation of the Bayesian evidence and is similar in construction to the Akaike (\citeyear{Akaike:1974}) information criterion. We therefore decided to fix a number of our model parameters to avoid overfitting our data. For $M_{\mathrm{h,}0}$, we follow \cite{Conroy:2006}, who propose, based on their simulations, the relationship 
\begin{equation}
\log_{10}M_{\mathrm{h},0}=0.76\log_{10}M_{\mathrm{h,}1}+2.3\rm.     
\end{equation}
We also choose to fix $\sigma_{\log M_{\rm h}}=0.2$ and $\alpha_{\rm sat}=1.0$ as these are standard values adopted throughout the literature and are supported by both theoretical and observational studies \citep[see e.g.][and references therein]{Wake:2011,Martinez-Manso:2015,Harikane:2016}.  There are thus three free parameters in our model: $M_{\mathrm{h,min}}$, $M_{\mathrm{h,}1}$ and $\Delta_{z}$.  We discuss our decision to leave $\Delta_{z}$ as a free parameter in more detail in Appendix~\ref{sec:delta_z_discuss}.  
  
We simultaneously fit both the observed clustering and number of galaxies by summing both contributions to the total $\chi^{2}$ such that
\begin{multline}
\chi^{2} = \sum_{i,j}[w^{\rm obs}(\theta_{i})-w^{\rm mod}(\theta_{i})](C^{-1})_{i,j}[w^{\rm obs}(\theta_{j})-w^{\rm mod}(\theta_{j})] \\ + \frac{[N_{\rm gal}^{\rm obs} - n_{\rm gal}^{\rm mod}\,V^{\rm mod}]^{2}}{\sigma^{2}_{\sqrt{N}} + \sigma^{2}_{\rm CV} + \sigma^{2}_{\mathrm{fit}}}\rm,
\end{multline}
where $n_{\rm gal}^{\rm mod}$ is the number density of galaxies predicted by the HOD and the volume, $V^{\rm mod}$, is computed according to $V^{\rm mod}=(\Omega/4\pi)\int W(z)\,(\mathrm{d}V/\mathrm{d}z)\,\mathrm{d}z$ where $\Omega$ is the area of our survey and $W(z)$ is defined as in \autoref{eq:sigma_z_window}. This therefore incorporates the effect of photometric redshift error on the observed number of galaxies through the $\Delta_{z}$ parameter. There are contributions to the error on the observed number of galaxies from Poisson noise, $\sigma_{\sqrt{N}}$, cosmic variance, $\sigma_{\rm CV}$, which is computed using the method presented in \cite{Moster:2011}, and a term that accounts for the error in the SED fitting procedure, $\sigma_{\rm fit}$.  For this latter term we construct $100$ mock catalogs by scattering each SMUVS galaxy within the probability distribution for its stellar mass and redshift given by \textsc{LePhare} (the construction of these mock catalogs are described in more detail by Deshmukh et al. \citeyear{Deshmukh:2018}). We then apply the same photometric redshift and stellar mass cuts to these mock catalogs and take $\sigma_{\rm fit}$ to be the standard deviation in the number of galaxies in each in each sample across the 100 mock catalogs. The value of these different sources of error for each sample is given in Table~\ref{table:main_results}, the total error on the observed number of galaxies is typically $\sim10$~percent. This is comparable to the uncertainty quoted by \cite{Wake:2011}, who derived errors of $\sim15$~percent on their galaxy number densities using a different method for a slightly smaller area than surveyed here ($\sim0.4$~deg$^{2}$). Additionally, we have scaled the inverse covariance matrix, $C^{-1}$, (see \autoref{eq:covar_matrix}) in order to account for bias from a finite number of jackknife samples, according to \cite{Hartlap:2007}. 

We assume uninformative (i.e., flat) priors for our three free parameters (our fitting procedure is thus analogous to a maximum likelihood estimation). After a conservative `burn-in' phase we run $40$~walkers for $2500$ loops, which results in the posterior distribution being sampled with $10^5$ points. The chains appear `well-mixed' by this point, and inspection of their autocorrelation and the Gelman-Rubin statistic \citep{GelmanRubin:1992} indicates that they have converged. Our best-fit parameters are taken to be the sample in our chains that returns the minimum $\chi^{2}$, and the uncertainties represent the bounds of the $1\sigma$ contours in parameter space described by $\chi^{2}<\chi^{2}_{\rm min}+\Delta\chi^{2}$, where $\Delta\chi^{2}=3.53$ for three free parameters. The value and uncertainties of the derived quantities (e.g., satellite fraction) are determined in the same way. Some examples of the likelihood distributions produced by our modeling and fitting procedure are shown in \autoref{fig:corner_example}, Appendix~\ref{sec:likelihood_example}. 
\section{Results}
In this section we display our main results. In Section~\ref{sec:halo_model_results} we present some of our measured angular correlation functions and the resulting best-fit halo models. We present the characteristic halo masses of, and satellite galaxy fractions derived from, our best-fit halo models in Sections~\ref{sec:char_halo_mass_results} and \ref{sec:fsat} respectively. In Section~\ref{sec:corr_length_bias_results} we show the clustering amplitude (comoving correlation length and large-scale bias) evolution inferred from our data, and in Section~\ref{sec:evolution} we use our bias measurements to propose a technique for computing coevolutionary histories for dark matter and stellar mass assembly. Finally, in Section~\ref{sec:shmr} we investigate the SHMR of our data. We show that with our derived SHMRs we can reproduce our stellar mass functions in Appendix~\ref{sec:smf}, highlighting the consistency of our analysis. Our main results are summarized in \autoref{table:main_results} in Appendix~\ref{sec:main_results_table}.    
\label{sec:results}
\subsection{Halo model analysis of angular two-point correlation functions}
\label{sec:halo_model_results}
\begin{figure*}
\centering
\includegraphics[width = 8.6189cm]{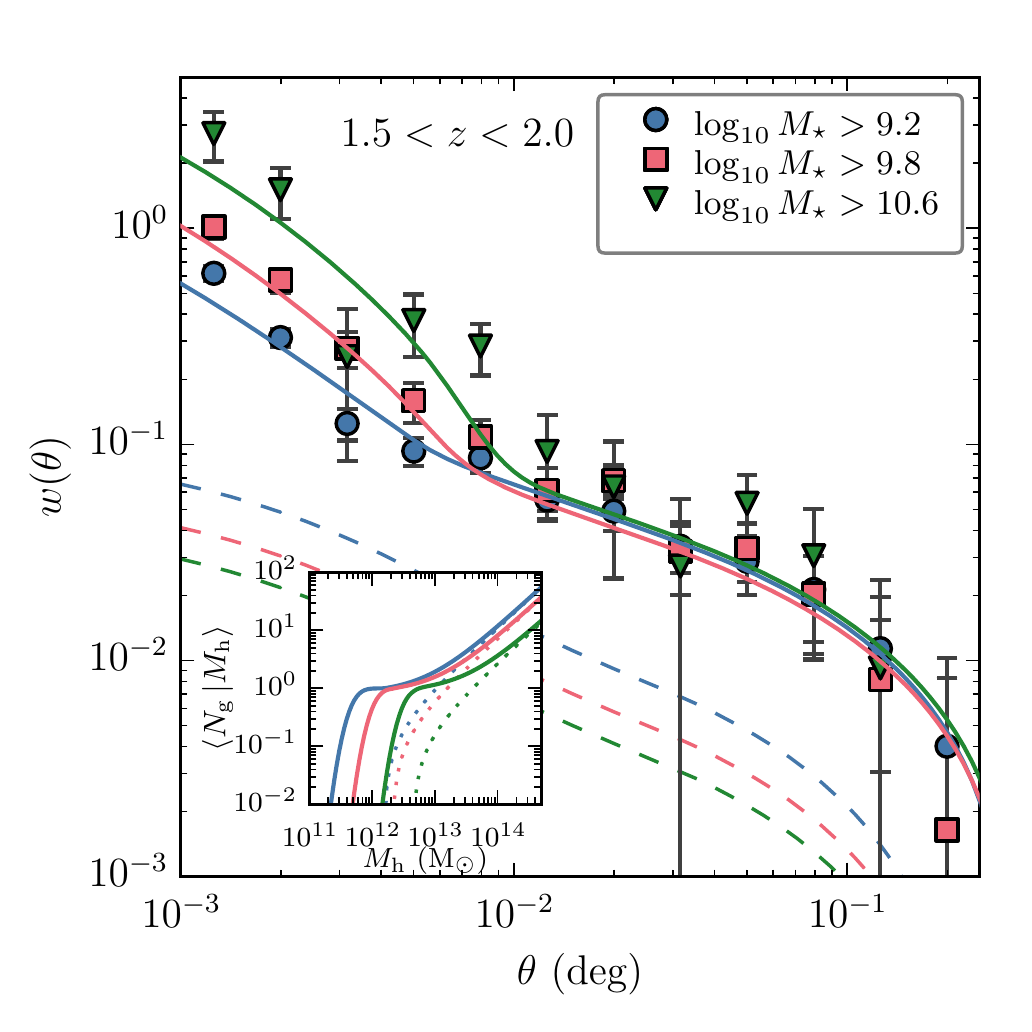}\hspace{0.79173cm}\includegraphics[width = 8.6189cm]{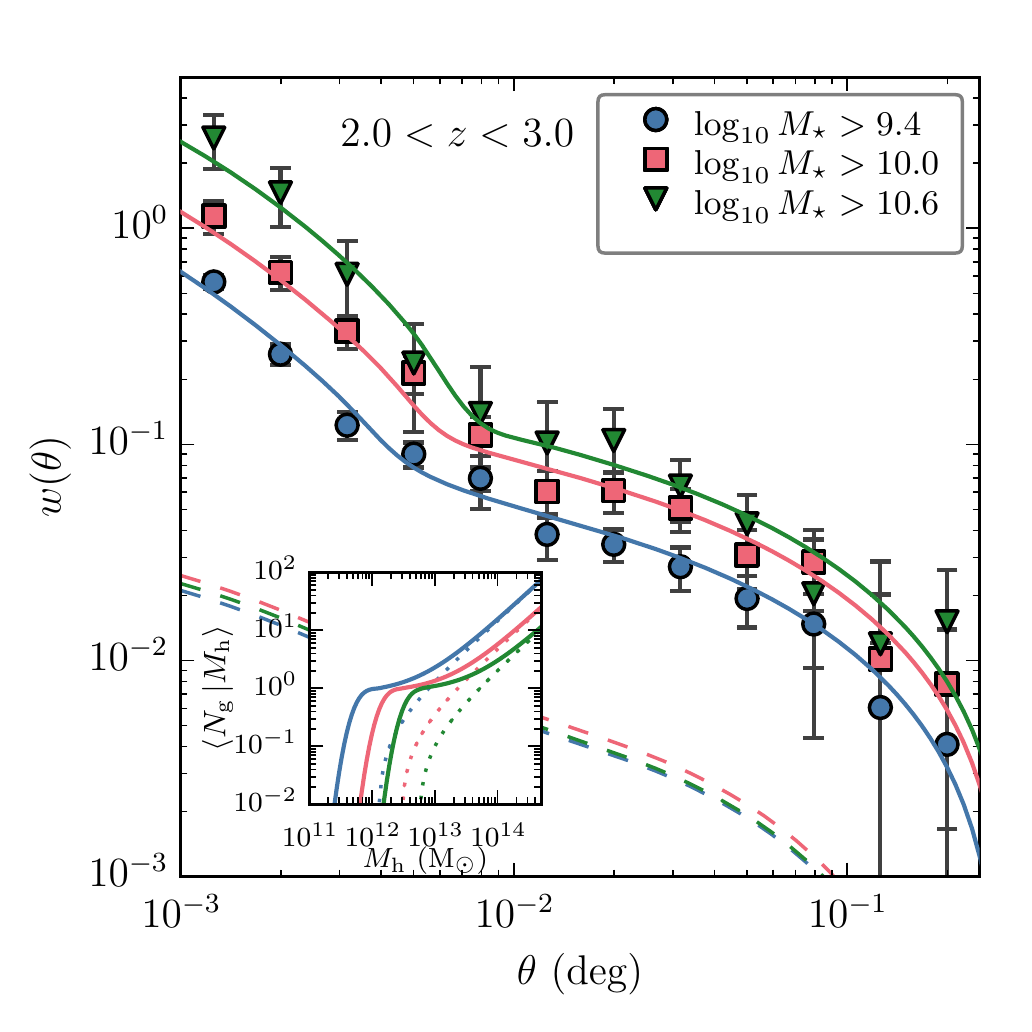}\\
\includegraphics[width = 8.6189cm]{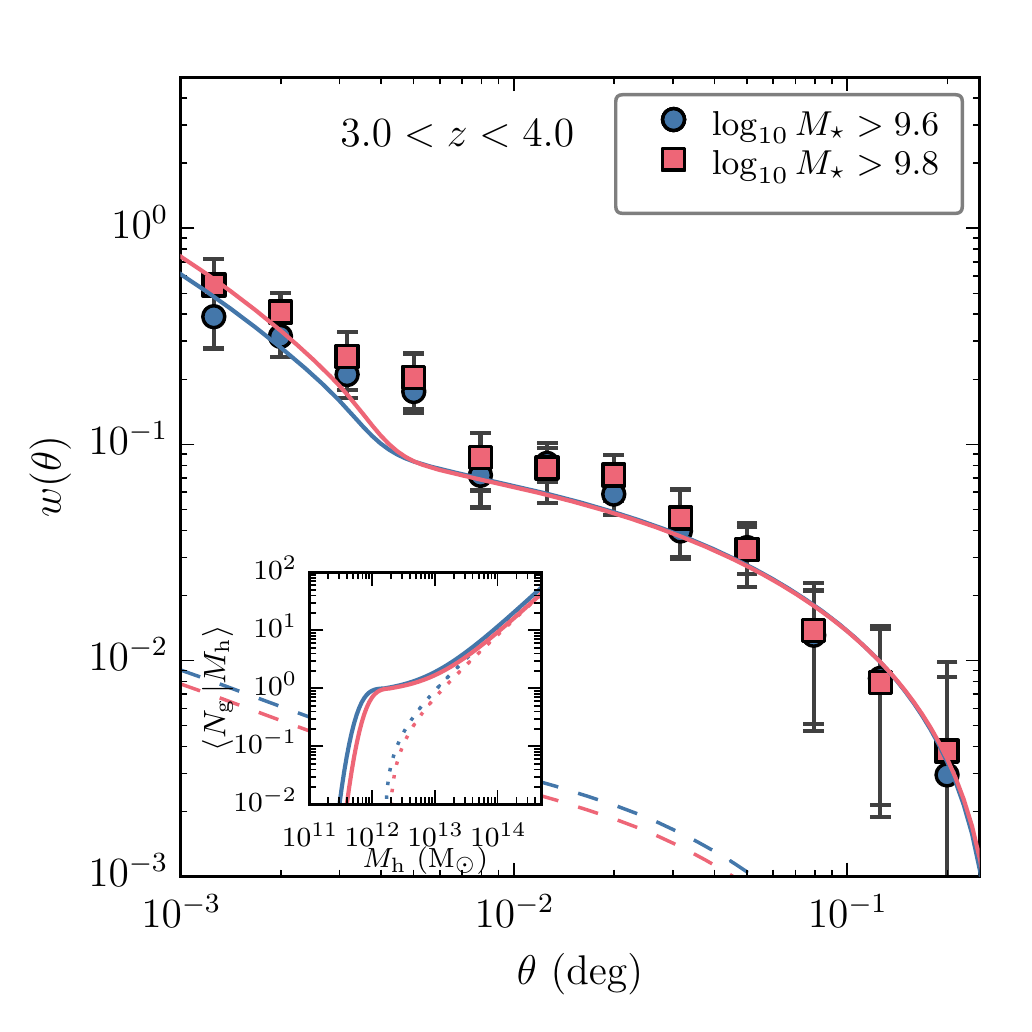}\hspace{0.79173cm}\includegraphics[width = 8.6189cm]{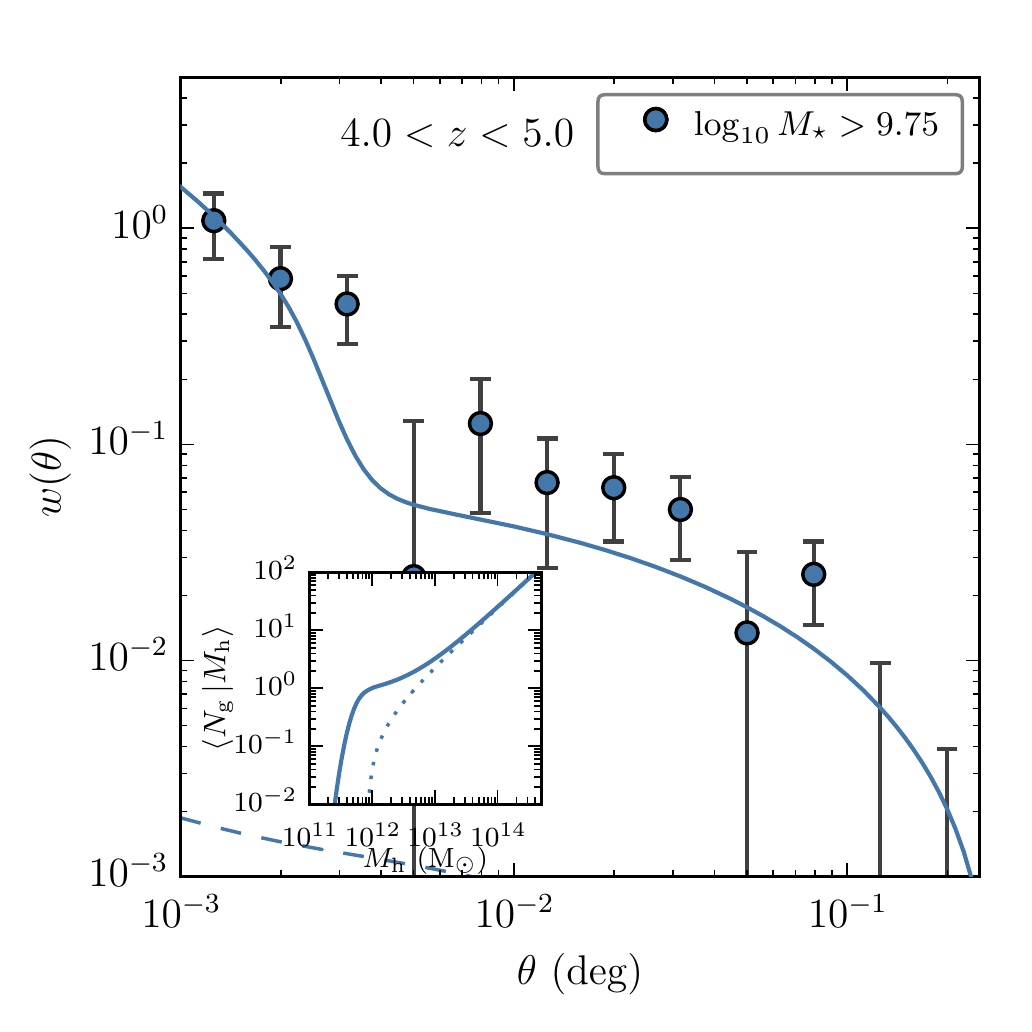}
\caption{Stellar mass-selected angular correlation function measurements in SMUVS. Each panel represents a different redshift bin. Colors and symbols indicate different stellar mass limits as shown in the legends. Solid lines show the angular correlation function of our best-fit halo models. The dashed lines indicate the angular correlation function of dark matter; note that these have been projected incorporating the best-fit value for $\Delta_{z}$. The inset panels show the best-fit halo occupation distribution; dotted lines here show the contribution from satellite galaxies.}
\label{fig:wtheta_hod}
\end{figure*}
Here we present our measurements of the clustering of galaxies and results from the corresponding best-fit halo model. Some examples of these are shown in \autoref{fig:wtheta_hod}.

In general, the fits to the observed clustering appear to be good. However, a purely visual inspection of the fit may be misleading. This is because the covariance between different angular bins and the abundance constraint used in the fitting are not visualized in \autoref{fig:wtheta_hod}. Our goodness of fit is also reflected in our low values for the minimum reduced $\chi^{2}$ (typically $\sim1$, see \autoref{table:main_results}).

It should be noted that the amplitudes of the measured angular correlation functions do not necessarily display a monotonically increasing trend with increasing stellar mass, evident in the $1.5<z<2.0$ panel of \autoref{fig:wtheta_hod}. However, as higher stellar mass galaxies are less abundant one would expect them to reside in higher mass halos\footnote{This is in fact constrained to be the case by the assumption that our central galaxy HOD reaches an amplitude of unity, the fixed value of $\sigma_{\log M_{\rm h}}$ and that the observed number of galaxies is used as a constraint in our fitting procedure.} that are more biased and would thus have a greater correlation amplitude. 

In our framework, this is accommodated by the $\Delta_{z}$ parameter in our fitting procedure. For example, at $1.5<z<2.0$ the $\log_{10}(M_{\star}/\mathrm{M}_{\odot})>9.8$ sample has a lower clustering amplitude on large scales ($\gtrsim5\times10^{-3}$~deg) than the $\log_{10}(M_{\star}/\mathrm{M}_{\odot})>9.2$ sample (see the top left panel of \autoref{fig:wtheta_hod}). Therefore, the higher mass sample must have a higher best-fit value of $\Delta_{z}$.  Our modeling produces a value of $\Delta_{z}\sim0.09$ relative to $\Delta_{z}\sim0.04$ for the high- and low-mass samples mentioned above respectively. We caution against overinterpreting these values in terms of $|z_{\rm phot}-z_{\rm spec}|/(1+z_{\rm spec})$ diagnostics, as our $\Delta_{z}$ parameter describes a global photometric redshift dispersion. It does not segregate out redshift outliers and ignores selection effects introduced by comparing to spectroscopic redshifts. A higher value of $\Delta_{z}$ projects the intrinsic spatial correlation function over a larger redshift range (according to \autoref{eq:limber}) and thus lowers the amplitude of the angular correlation function. This is also reflected in the amplitudes of the matter correlation functions, shown as the dashed lines in \autoref{fig:wtheta_hod}.

Generally, though it is not always the case, we find that the samples with higher stellar mass in a given redshift bin tend to have higher values of $\Delta_{z}$. This is not necessarily unexpected. Massive galaxies ($M_{\star}\gtrsim10^{10}$~M$_{\odot}$) exhibit, on average, greater dust attenuation than less massive ones. This, in turn, leads to an increased degeneracy in the parameter space of the SED fitting between stellar age and dust reddening\footnote{Interestingly, our values for $\Delta_{z}$ are lower for $2.0<z<3.0$ than $1.5<z<2.0$, as our redshift bin is broader and the constraint of the age of the Universe (and thus of the stellar populations) at these higher redshifts reduces this degeneracy \citep[e.g.][]{Thomas:2017}.}. In general, however, we find that our best-fit values of $\Delta_{z}$ are low relative to the size of our photometric redshift bins, suggesting that our redshifts are accurate enough for the analysis performed in this study. Typically, we find $\Delta_{z}(1+z_{50})/(z_{\rm hi}-z_{\rm lo})\lesssim0.5$ (where $z_{50}$ is the median redshift of the sample and $z_{\rm hi}$ and $z_{\rm lo}$ represent the limits of the bin), and all of our values for $\Delta_{z}(1+z_{50})/(z_{\rm hi}-z_{\rm lo})$ are consistent with being lower than unity (within $1\sigma$, see \autoref{fig:delta_z_abs} and the discussion in Appendix~\ref{sec:delta_z_discuss} for more details). 
\subsection{Characteristic halo masses}
\label{sec:char_halo_mass_results}
\begin{figure}
\centering
\includegraphics[width = 8.6189cm]{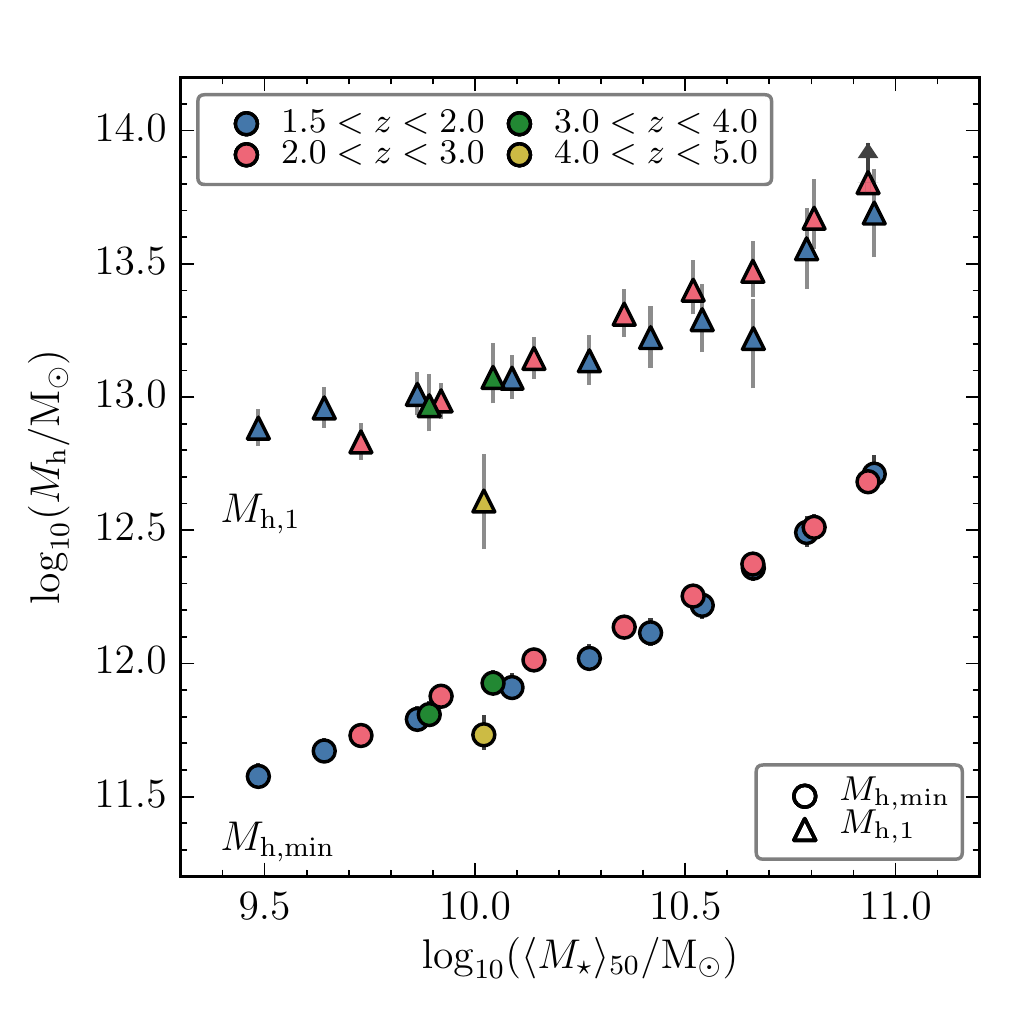}
\caption{Characteristic halo masses $M_{\rm h,min}$ (circles) and $M_{\rm h,1}$ (triangles) as a function of galaxy sample median stellar mass. Colors indicate different redshift bins, as shown in the upper legend.}
\label{fig:halo_masses}
\end{figure}
Here we consider the characteristic halo masses, $M_{\rm h,min}$ and $M_{\rm h,1}$, which represent the mass at which $50$~percent of halos host a single central galaxy in the sample, and at which each halo hosts an additional satellite galaxy respectively\footnote{In fact, the halo mass at which the HOD is equal to $2$ is not exactly $M_{\rm h,1}$, due to the cutoff halo mass, $M_{\mathrm{h,}0}$, and the scaling of the satellite galaxy distribution by that of the central galaxy distribution. However, for our purposes here this distinction is unnecessary.}. These quantities are shown in \autoref{fig:halo_masses} as a function of the median stellar mass of each sample, and redshift. 

We can see that both halo masses are well constrained by our clustering and abundance measurements and that they form tight, approximately linear, relationships with stellar mass (in log space) with no significant evolution with redshift (though our $4.0<z<5.0$ point is offset to lower halo masses, as is expected from the hierarchical nature of structure formation). We find the `mass gap' between $M_{\rm h,min}$ and $M_{\rm h,1}$ to be $\sim10-20$ which is broadly consistent with earlier studies \citep[e.g.,][]{Zehavi:2011,McCracken:2015}. 
\subsection{Satellite fractions}
\label{sec:fsat}
\begin{figure}
\centering
\includegraphics[width = 8.6189cm]{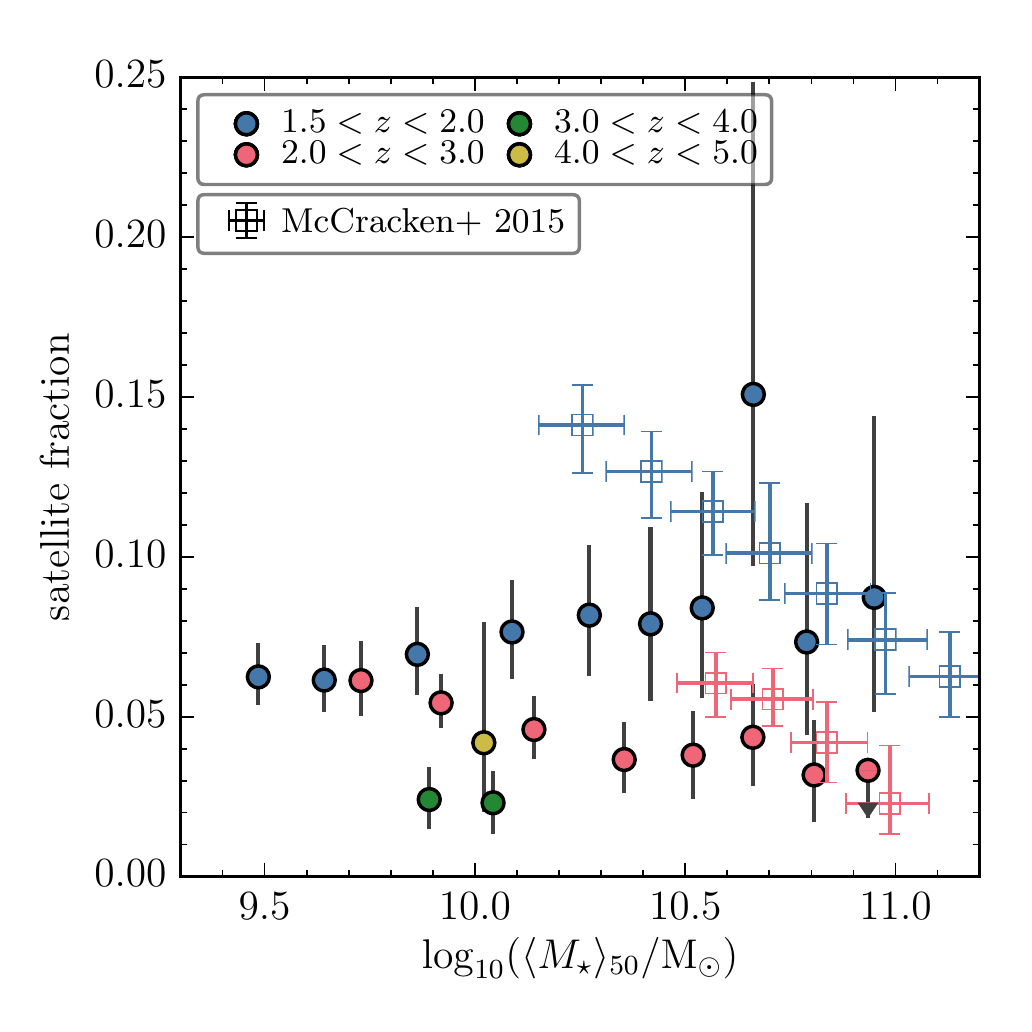}
\caption{Satellite fraction as a function of galaxy sample median stellar mass. Colors indicate different redshift bins, as shown in the legend. Observational data from McCracken et al. (\citeyear{McCracken:2015}, open squares) are also shown.}
\label{fig:fsat}
\end{figure}
Satellite galaxies are those that do not sit at the center of the potential well of a dark matter halo (as `central' galaxies do), but rather formed in other dark matter halos that later merged with (or were accreted onto) their current host. They are on bound orbits, which decay due to the dynamical friction of dark matter acting on the sub-halo, and will eventually merge onto the central galaxy. Thus an increase in the satellite fraction can indicate halo merging activity, whilst a decrease can indicate satellites merging with the central galaxy. 

The satellite fractions derived from our clustering and abundance measurements (\autoref{eq:fsat}) are shown in \autoref{fig:fsat}. We find that our inferred satellite fractions generally increase with cosmic time, as can be expected from hierarchical structure formation. This is supported by studies at lower redshift that indicate higher satellite fractions than we observe here, for example, \cite{Zehavi:2011} derive a value of $f_{\rm sat}\approx0.3$ at $z=0$. Additionally, we observe a mild decreasing trend of $f_{\rm sat}$ with increasing stellar mass for $2.0<z<3.0$, but a much flatter relationship for $1.5<z<2.0$. This is indicative of halo merging activity (which converts central galaxies into satellites) and/or in-situ stellar mass assembly in satellite galaxies (which moves satellite galaxies into higher stellar mass samples). However, distinguishing the relative contribution of these two potential processes would require further analysis of these data with more physically motivated models than the HOD framework adopted here.

Our data compare favorably with the satellite fractions inferred by \cite{McCracken:2015}.  These authors performed a similar halo model analysis as performed in this work, though their initial photometry was based on the earlier UltraVISTA DR1 \citep{McCracken:2012}. Additionally, they left all of the HOD parameters as free in their fitting procedure (whereas some are fixed here as was discussed in Section~\ref{sec:fitting}) and did not consider photometric redshift dispersion in their model, equivalent to assuming $\Delta_{z}=0$ in this work. For $2.0<z<3.0$ the two studies are in excellent agreement, though it should be noted that the McCracken et al. redshift range for $z\gtrsim2$ spans only $2.0<z<2.5$. However, the agreement is less good for $1.5<z<2.0$ and we do not reproduce the decreasing trend with increasing stellar mass found by McCracken et al. (and evident to a lesser extent in our measurements for $2.0<z<3.0$) and other studies at lower redshift \citep[e.g.][]{Coupon:2012}. 

Our satellite fraction, and those of McCracken et al., are smaller than those found by \cite{Wake:2011} and \cite{Martinez-Manso:2015}, who derived satellite fractions in the range $f_{\rm sat}~0.1-0.25$ at $z\sim1.5-2$ for $\log_{10}(M_{\star}/\mathrm{M}_{\odot})>10$ galaxies. The Wake et al. study is based on data from the NEWFIRM medium-band survey \citep[NMBS,][]{vanDokkum:2009} and is drawn from $\sim0.4$~deg$^{2}$, suggesting that the larger area of the McCracken et al. study ($\sim1.5$~deg$^2$) and ours ($\sim0.66$~deg$^2$) may be responsible for the difference. However, the Martinez-Manso et al. study is based on the $\sim95$~deg$^{2}$ \emph{Spitzer} South Pole Telescope Deep-Field Survey \citep{Ashby:2013}. The satellite fraction is most sensitive to the HOD parameters $\alpha_{\rm sat}$ and $M_{\mathrm{h,}1}$ which are constrained primarily by the small-scale ($\theta\lesssim10^{-2}$~deg) clustering relating to the one-halo term in the HOD model. It is unlikely that our choice to fix $\alpha_{\rm sat}$ is the cause, as both the Wake et al. and Martinez-Manso et al. studies also fixed $\alpha_{\rm sat}=1$. Therefore it may be the case that some features in the small-scale clustering in the COSMOS field may result in lower satellite fractions in our work and the McCracken et al. study than Wake et al. and Martinez-Manso et al. (though Wake et al. drew half of their area from COSMOS). It should also be noted that the Martinez-Manso et al. study does not select galaxies by their stellar mass but instead by their mid-infrared color, which may complicate a direct comparison with this work.      
    
\subsection{Comoving correlation length and galaxy bias}
\label{sec:corr_length_bias_results}
\begin{figure}[h]
\centering
\includegraphics[width = 8.6189cm]{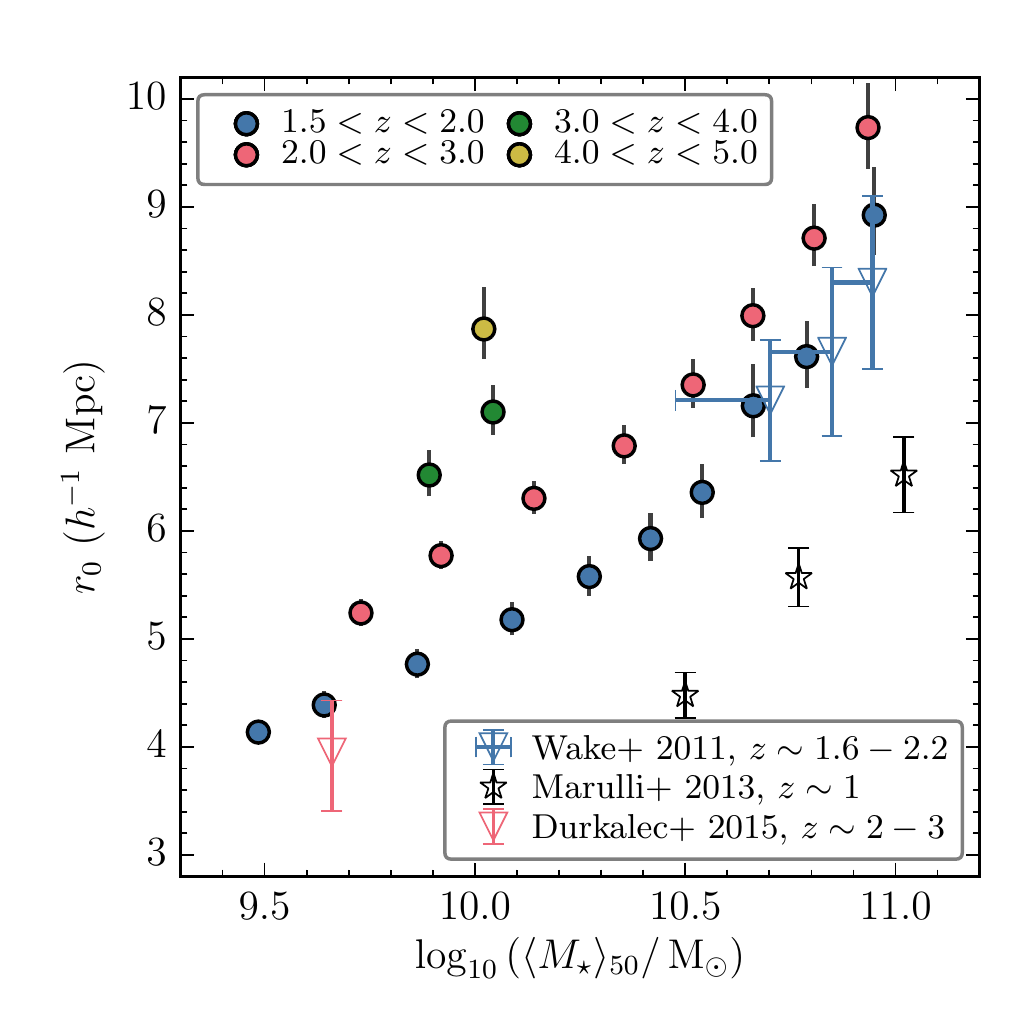}
\caption{Comoving correlation length as a function of galaxy sample median stellar mass. Colors indicate different redshift bins, as shown in the legend. Observational data from Wake et al. (\citeyear{Wake:2011}, open blue triangles), Marulli et al. (\citeyear{Marulli:2013}, open stars), and Durkalec et al. (\citeyear{Durkalec:2015}, open red triangle) are also shown. We have approximately scaled the Wake et al. data from their quoted stellar mass threshold to a median stellar mass by applying a similar shift between these two quantities as found in the SMUVS data. This change is indicated by the horizontal error bars on the Wake et al. data.}
\label{fig:r0}
\end{figure}
\begin{figure*}
\centering
\includegraphics[trim = 0 0 0 0, clip = True, width = 8.6189cm]{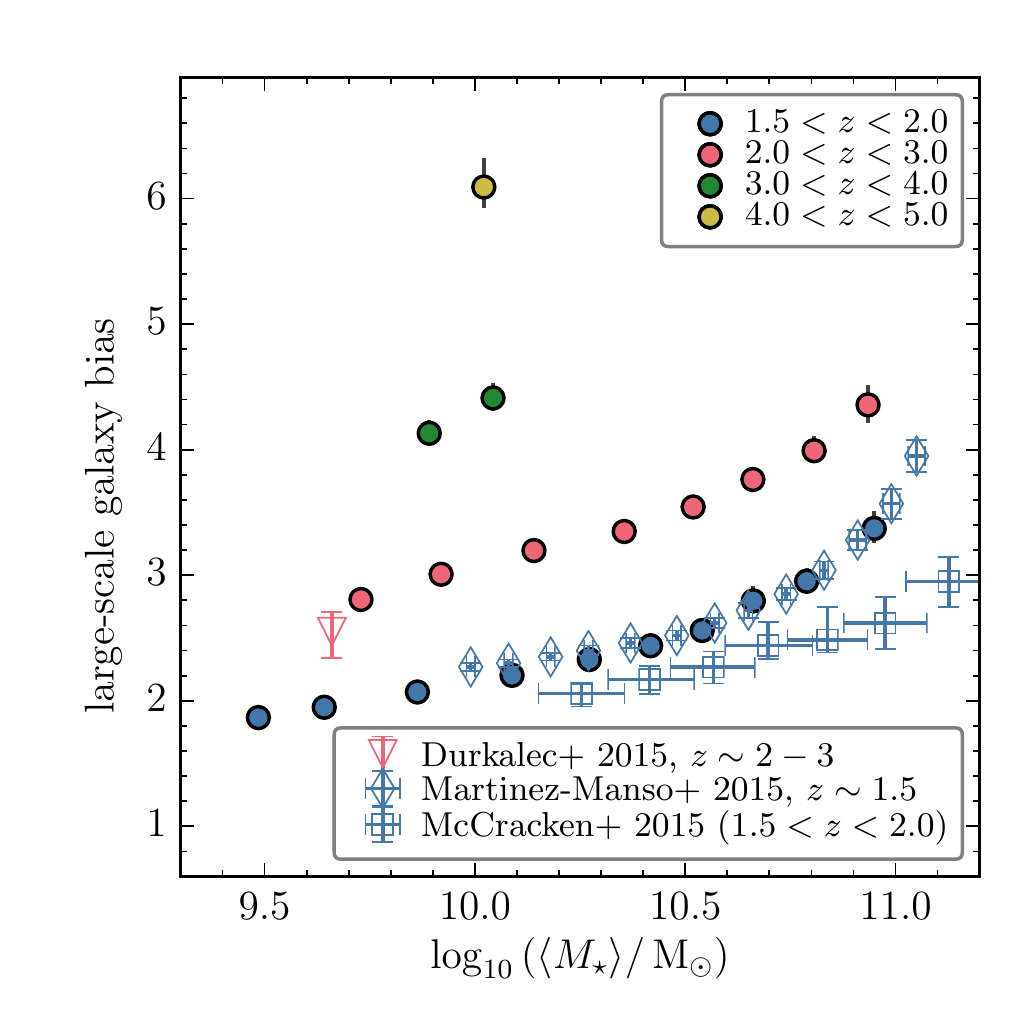}\hspace{0.79173cm}\includegraphics[trim = 0 0 0 0, clip = True, width = 8.6189cm]{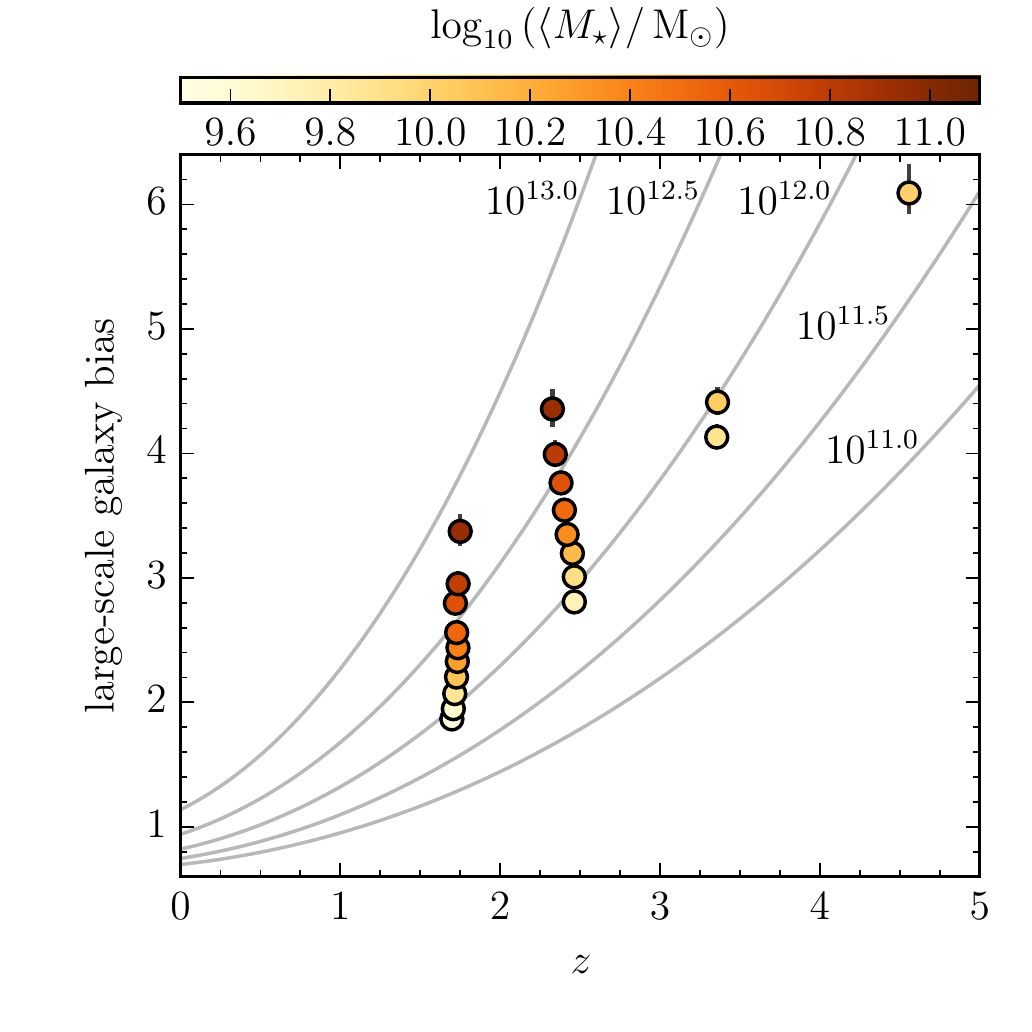} 
\caption{Inferred large-scale galaxy bias as a function of stellar mass and redshift. \emph{Left panel}:  large-scale bias as a function of galaxy sample median stellar mass. Colors indicate different redshift bins, as shown in the legend. Other observational data are from Durkalec et al. (\citeyear{Durkalec:2015}, open triangle), Martinez-Manso et al. (\citeyear{Martinez-Manso:2015}, open diamonds), and McCracken et al. (\citeyear{McCracken:2015}, open squares) are also shown. \emph{Right panel}:  large-scale bias as a function of redshift. The color scale indicates the galaxy sample median stellar mass. Solid gray lines indicate the bias evolution of halos of a constant mass (indicated in the panel), according to the prescription of \cite{Tinker:2010}.}
\label{fig:large_scale_bias}
\end{figure*}

Earlier studies of the clustering of galaxies at low redshift found that their correlation function could be adequately described by a power law, i.e., $\xi(r) = (r/r_{0})^{-\gamma}$ \citep[e.g.,][]{DavisPeebles:1983,Hawkins:2003}. Here $r_{0}$ is the `correlation length' that describes the separation at which the correlation function is equal to unity, and $\gamma$, which is typically $\sim1.8$, describes the slope. 

However, this description of the correlation function of galaxies is somewhat problematic as more recent observations, at higher redshifts, have shown that galaxy clustering exhibits deviations from a simple power law that can be readily understood in terms of halo occupation models such as those used in this work \citep[e.g.,][]{Zheng:2004}. Moreover, the results from fitting a simple power law can be sensitive to the scales used in the fitting procedure and whether $\gamma$ was fixed or not. However, it can still be informative, both for comparing with earlier work and as it roughly describes the amplitude of the observed correlation function. 

Here, rather than attempt to fit a power law to our observed angular correlation functions, we compute this quantity directly from the spatial correlation functions predicted by our best-fit halo models. For this we define the correlation length such that $\xi(r_{0})\equiv1$. We show our inferred $r_{0}$ values in \autoref{fig:r0}. Here, we can see that $r_{0}$ generally increases monotonically with redshift and stellar mass. For example, at $1.5<z<2.0$ galaxies with $M_{\star}\sim10^{10.3}$~M$_{\sun}$ display a correlation length of $\sim5.6$~$h^{-1}$~Mpc, whereas at $4.0<z<5.0$ the correlation length for galaxies of a similar stellar mass is $\sim7.9$~$h^{-1}$~Mpc.

Comparing to some earlier measurements in \autoref{fig:r0} we can see that our data occupy a hitherto relatively unexplored region of this parameter space. \cite{Marulli:2013} fit a power law to the projected spatial correlation function measured from galaxies identified in the VIMOS public extragalactic redshift survey \citep[VIPERS, ][]{Guzzo:2013} and \cite{Durkalec:2015} performed a similar analysis with data from the VIMOS ultra-deep survey (VUDS). Both of these VIMOS surveys allow for the determination of spectroscopic redshifts. The Marulli et al. data appear broadly consistent with the trends identified in ours, though the redshift range of the two studies does not overlap. They find the same trend of increasing $r_{0}$ with stellar mass and their correlation lengths are at smaller scales than ours (for a fixed stellar mass) which would be expected as they are measured at lower redshift. At lower stellar masses ($M_{\star}\sim10^{9.6}$~M$_{\odot}$) we find a larger $r_{0}$ than Durkalec et al., though this may be due their $r_{0}$ being determined through a power-law fit, rather than a halo model as we do here, as the large-scale bias Durkalec et al. inferred from fitting a HOD model to their clustering measurements agrees well with ours, as is shown in \autoref{fig:large_scale_bias}. \cite{Wake:2011} find similar correlation lengths to ours for high-mass galaxies, though they derive their correlation lengths by fitting a power law to their observed angular clustering, which is different to the procedure we follow.  

A perhaps more physically meaningful quantity, related to the comoving correlation length, is the large-scale galaxy bias. This describes the difference in amplitude between the matter correlation function and that of the observed galaxy sample. We compute an effective large-scale bias from our best-fit halo models, according to \autoref{eq:bias}. These are shown in \autoref{fig:large_scale_bias} as a function of stellar mass and redshift. A striking monotonically increasing relationship between redshift and large-scale bias (at a fixed stellar mass) is evident here. The relationship between stellar mass and bias appears to be flatter at lower redshifts ($z\lesssim3.0$), and at lower stellar masses (though we note that the stellar mass range probed also changes with redshift). 

A similar relationship between large-scale bias and stellar mass as is also seen at $z\sim1.5$ in the \cite{Martinez-Manso:2015} data, with which our determinations agree excellently. Our bias values are also in reasonable agreement with \cite{McCracken:2015}, though they find a flatter trend at high stellar masses.  This may be in part due to our inclusion of $\Delta_{z}$ as a free parameter, as calculations in which we fix $\Delta_{z}=0.035$ (as suggested by the $\sigma_{z}$ value for our catalog in the range $1.5<z<5.0$) produce a flatter relationship in better agreement with the McCracken et al. data. Martinez-Manso et al. account for the redshift dispersion as they use the full redshift probability distributions for each galaxy in projecting the spatial correlation function produced by their HOD model. The bias values found by McCracken et al. from their $2.0<z<2.5$ measurements ($b_{\rm gal}\lesssim3$) appear lower than those we infer for $2.0<z<3.0$, however, this may be due to the broader redshift range we consider, which would include more biased galaxies in the $2.5<z<3.0$ range (for a fixed stellar mass threshold) than the McCracken et al. data. We also note that the halo model-derived bias values found by \cite{Wake:2011} for $z\sim1.6-2.2$ are higher than ours for $1.5<z<2.0$. They find $b_{\rm gal}\sim3-4$ for galaxies with $\log_{10}(M_{\star}/\mathrm{M}_{\odot})>10.5$.    

Again, this comparison with literature values indicates that we have filled in a large hitherto unexplored area of this plane, and our results show reasonable agreement where they coincide with earlier measurements.       
\subsubsection{Potential evolutionary paths}
\label{sec:evolution}
\begin{figure}
\centering
\includegraphics[width = 8.6189cm]{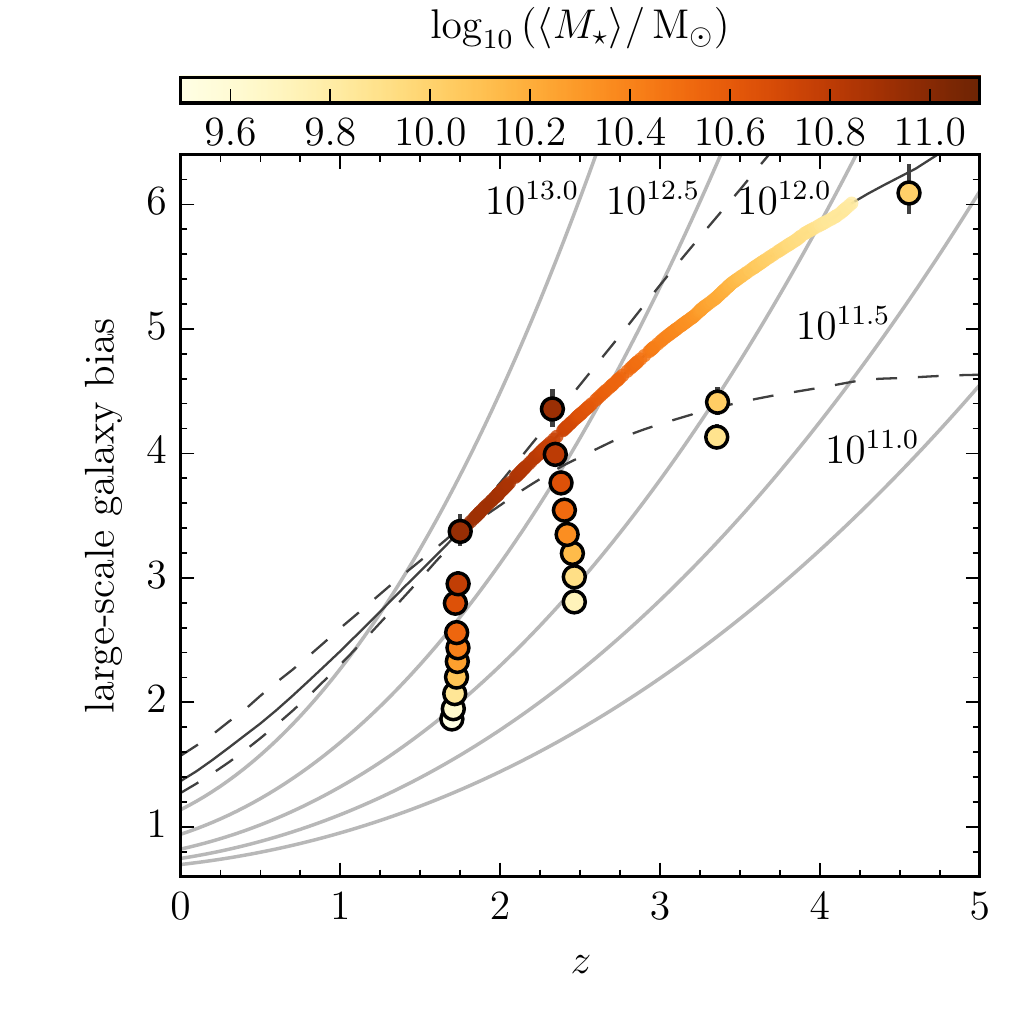}
\caption{Same as for the right panel of \autoref{fig:large_scale_bias} but also indicating a potential evolutionary path for the host dark matter halos and stellar mass content of the progenitors of galaxies with $\log_{10}(M_{\star}/\mathrm{M}_{\odot})>10.8$ at $1.5<z<2.0$. The solid black line indicates the bias of the median halo mass accretion history, which is computed according to \cite{McBride:2009}, and the colored line on top of this indicates the median stellar mass of the progenitors, computed following the cumulative abundance argument of \cite{Behroozi:2013a}. The dashed black lines indicate the $16-84$~percentile scatter of the halo mass accretion histories.}
\label{fig:bias_zevol}
\end{figure}
\begin{figure*}
\centering
\includegraphics[width = 8.6189cm]{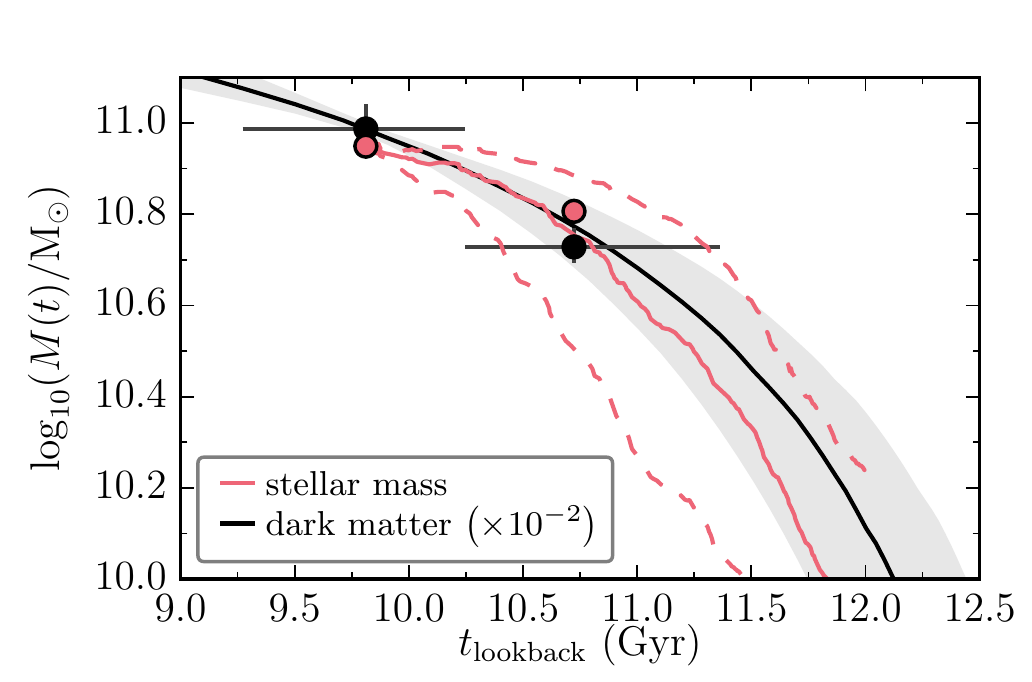}\hspace{0.79173cm}\includegraphics[width = 8.6189cm]{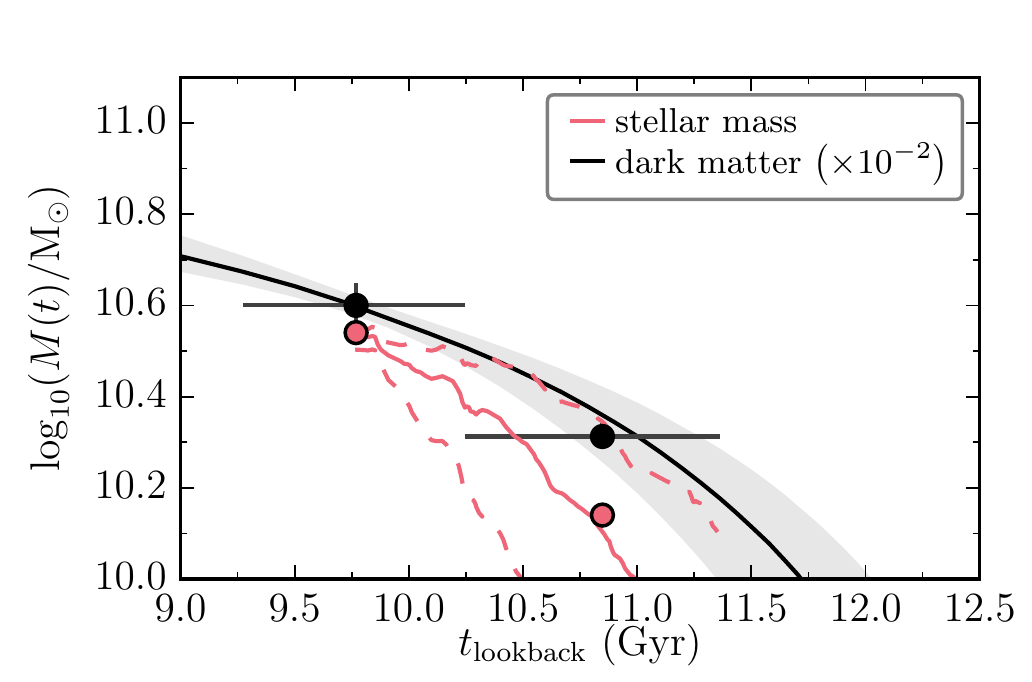}
\caption{Examples of our proposed evolutionary paths. Dark matter halo (black line) and stellar mass (red line) assembly histories computed as described in Section~\ref{sec:evolution}, as a function of lookback time. Halo masses are converted from bias measurements presented in \autoref{fig:bias_zevol} using the relation of \cite{Tinker:2010}. The gray shaded region describes the $16-84$~percentile scatter of the halo mass accretion histories. The median of dark matter accretion histories has been scaled by $10^{-2}$ for presentation purposes. Data from \autoref{fig:bias_zevol} that intersect this history are also shown as points with error bars. The horizontal error bars indicate the range of the redshift bin that point is drawn from, in terms of lookback time. The dashed lines indicate the $1\sigma$ scatter on the stellar mass history. The starting point for the histories calculation is galaxies with $\log_{10}(M_{\star}/\mathrm{M}_{\odot})>10.8$ in the left panel and $>10.2$ in the right panel.}
\label{fig:coevol}
\end{figure*}

In this section, we propose a method to track the coevolution of dark matter halo and stellar mass assembly, based on our large-scale bias and stellar mass measurements presented above. 

First, we generate descendant halo masses, according to our $z=0$ halo mass function. To these we apply halo mass assembly histories, using the form proposed by \cite{McBride:2009}. These authors parametrized halo mass accretion histories from the Millennium $N$-body simulations \citep{Springel:2005} according to
\begin{equation}
M_{\rm h}(z) = M_{\mathrm{h,}z=0}\,(1+z)^{\beta}\,e^{-\gamma z}\rm,
\end{equation}
where $M_{\mathrm{h,}z=0}$ is the mass of the $z=0$ descendant. We note that other parametrizations for halo mass accretion histories in $\Lambda$CDM have been proposed in the literature \citep[e.g.,][]{vdBosch:2002,Wechsler:2002}. We generate a realistic ensemble of these histories according to the distributions of $\beta$ and $\gamma$ described in Appendix A of McBride et al. We then select halo histories with a bias equal to the effective galaxy bias we measure for a given galaxy population at a given redshift and track the evolution of this bias over cosmic time.  We use the halo mass--large-scale bias relation of \cite{Tinker:2010} to convert halo mass into large-scale bias in this section, in order to be consistent with our earlier analysis. 

Using our galaxy population with $\log_{10}(M_{\star}/\mathrm{M}_{\odot})>10.8$ at $1.5<z<2.0$ as our starting point, we show the median evolution of the bias for such histories as the solid dark line in \autoref{fig:bias_zevol}. The dashed lines indicate the $16-84$~percentile scatter which we can compute from the ($\sim10^3$) halo histories that go through this locus on the bias-redshift plane. At high redshift, the scatter in the halo histories encompasses a broad range of progenitor halo masses.  

For stellar mass evolution, we apply the cumulative number density approach proposed by \cite{Behroozi:2013a}. These authors found, using the abundance model of \cite{Behroozi:2013b}, that tracking the progenitors of galaxies selected by their stellar mass at a given redshift could be done simply by increasing the cumulative number density of the population by $0.16 \times\Delta z$~dex at previous redshifts\footnote{In practice, we used the Behroozi et al.\ code publicly available at \url{https://code.google.com/archive/p/nd-redshift/} to compute the progenitor number density and its $1\sigma$ scatter.}. Behroozi et al. attributed this simple power-law behavior to the roughly constant halo merger rate per unit halo per unit $\Delta z$ in $\Lambda$CDM cosmologies \citep[e.g.,][]{Fakhouri:2010}. 

We do this now for our SMUVS data, starting again with $\log_{10}(M_{\star}/\mathrm{M}_{\odot})>10.8$ galaxies at $1.5<z<2.0$. At earlier redshifts, we relate the progenitor number density (based on the Behroozi et al. argument outlined above) to a stellar mass threshold and a median stellar mass. The evolution of the median stellar mass is shown as the colored line, plotted on top of the median halo mass accretion history described above, in \autoref{fig:bias_zevol}. We track the stellar mass evolution to $z\sim4$, at which point the stellar mass threshold of the progenitors falls below our $80$~percent stellar mass completeness limit.

Of interest is where our median halo history intersects with our measurements of galaxy bias at higher redshifts. Here, our median stellar masses computed via the progenitor number density method outlined above appear to agree broadly with those measured directly at that redshift for galaxy samples with a similar bias to our computed median halo mass accretion history. 

This is a potentially powerful result, as it indicates that clustering measurements of stellar mass-selected galaxy samples, such as those performed here, can be used to project consistent evolutionary paths for the dark matter and stellar mass assembly over significant proportions of the history of the Universe.  It is also more straightforward to compute than other methods used to propose evolutionary pathways for galaxies based on clustering measurements \citep[e.g.,][]{Conroy:2008}.  We present this result again in the left panel of \autoref{fig:coevol}, where the assembly histories are instead plotted as a function of lookback time and the bias measurements from \autoref{fig:bias_zevol} have been converted into halo masses using the relation of \cite{Tinker:2010}. In the right panel of \autoref{fig:coevol} we show another example, applying this technique to galaxies with $\log_{10}(M_{\star}/\mathrm{M}_{\odot})>10.2$ at $1.5<z<2.0$. 

Also of note here is that the assembly histories for stellar mass and dark matter are intrinsically different shapes. This is a challenge for most physical galaxy formation models to reproduce, as is discussed in detail by \cite{Mitchell:2014}.               

There are a number of caveats to the result in this section, and it should be noted that the agreement between the proposed assembly histories and our data is not perfect and that the scatter on both the halo and stellar mass histories is significant. Choosing halo histories based on a single bias value ignores the fact that a range of halo masses will be occupied, which is indeed implied by our halo occupation models. This would probably only increase the (already significant) scatter and have a relatively minor impact on the median history, as there is a steep fall-off of the halo mass function toward high halo masses and in our HODs toward low halo masses. The halo histories we generate are based on simulations performed with slightly different cosmological parameters than those assumed here\footnote{The Millennium simulations assumed $\Omega_{\mathrm{m}}=0.25$, $\Omega_{\rm b}=0.045$, $\Omega_{\Lambda}=0.75$, $h=0.73$, $\sigma_{8}=0.9$, and $n_{\rm s}=1$.}, however, any change due to this is likely to be sub-dominant to the intrinsic scatter in the histories, which we have explicitly shown.  

Nevertheless, we consider the agreement good enough for this technique, combining $\Lambda$CDM halo mass accretion histories and cumulative number density arguments for stellar mass evolution, based on stellar mass-selected clustering measurements, to be explored further in order to aid our understanding of the coevolution of the stellar and dark matter content of halos.           
\subsection{The stellar to halo mass relation}
\label{sec:shmr}
\begin{figure*}
\centering
\includegraphics[width = \linewidth]{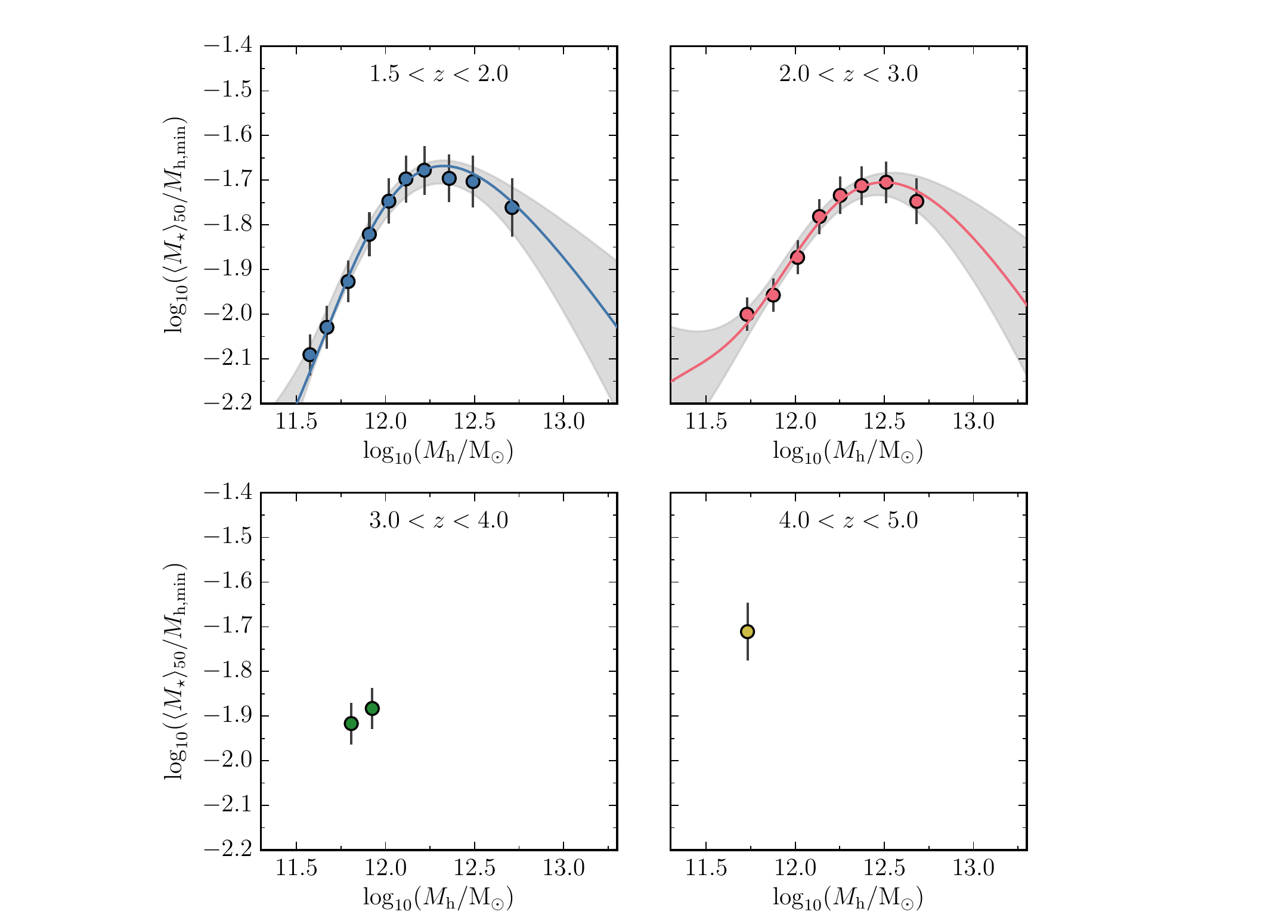}
\caption{Ratio between the median stellar mass and $M_{\rm h,min}$ for each galaxy sample for the redshifts indicated in the panel. The solid lines in the $1.5<z<2.0$ and $2.0<z<3.0$ panels indicate the best-fit SHMRs of \cite{Behroozi:2013b}, with the $1\sigma$ ($16-84$~percentile) uncertainties indicated by the gray shaded region.}
\label{fig:shmr}
\end{figure*}
\begin{figure}
\centering
\includegraphics[width = \columnwidth]{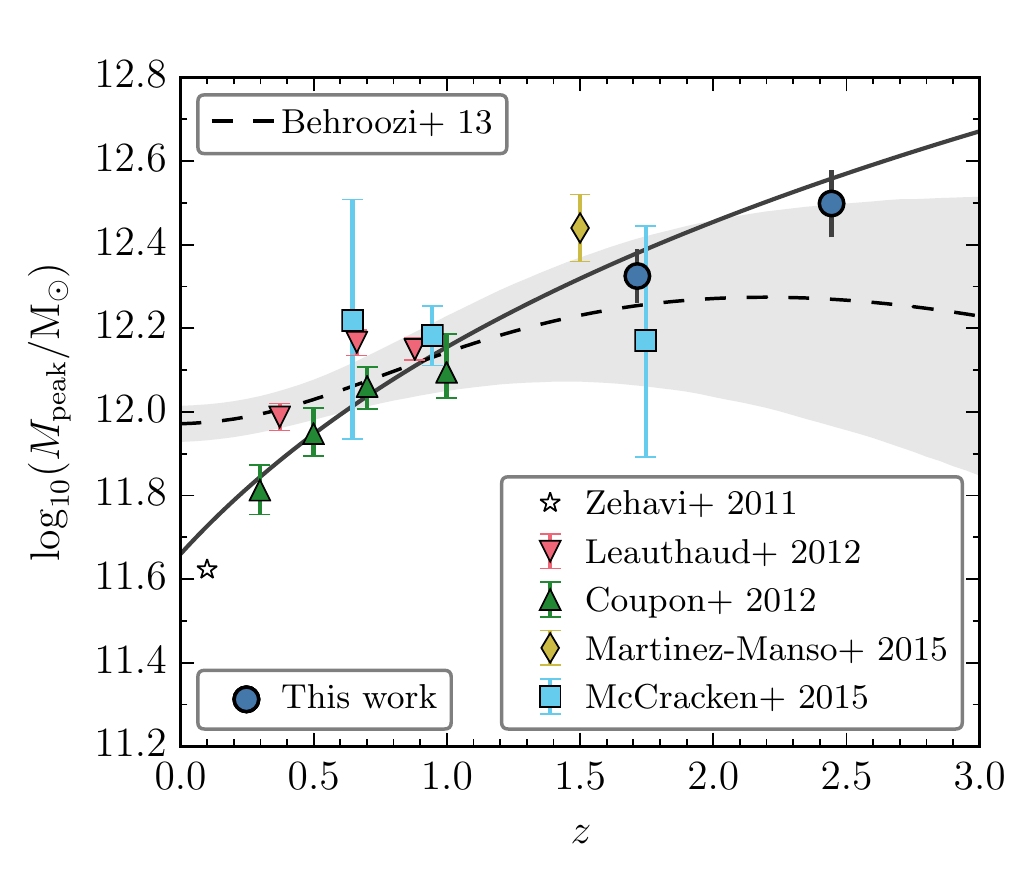}
\caption{Location of the maximum in the SHMR ($M_{\rm peak}$) as a function of redshift. Observational data from Zehavi et al. (\citeyear{Zehavi:2011}, open star), Leauthaud et al. (\citeyear{Leauthaud:2012}, red downward triangles), Coupon et al. (\citeyear{Coupon:2012}, green triangles), Martinez-Manso et al. (\citeyear{Martinez-Manso:2015}, yellow diamond), and McCracken et al. (\citeyear{McCracken:2015}, light blue squares) are also shown. The solid line indicates a best-fit power-law relation in $(1+z)$ to the observational data shown. Simulation data from the abundance-matching model of Behroozi et al. (\citeyear{Behroozi:2013b}, dashed black line) are also shown, with the gray shaded region indicating the $1\sigma$ uncertainty of this quantity.}
\label{fig:mpeak_z}
\end{figure}
\capstartfalse
\begin{deluxetable*}{ccccccc}
\tablecaption{SHMR best-fit parameter values}
\tablehead{
\colhead{redshift bin} & \colhead{$\log_{10}(\varepsilon)$} & \colhead{$\log_{10}(M_{\rm t}/\mathrm{M}_{\odot})$} & \colhead{$\alpha$} & \colhead{$\delta$} & \colhead{$\gamma$} & \colhead{$\log_{10}(M_{\rm peak}/\mathrm{M}_{\odot})$}}
\startdata
$1.5<z<2.0$&$-1.81_{-0.20}^{+0.10}$&$11.92_{-0.24}^{+0.20}$&$-1.64_{-0.31}^{+0.39}$&$3.30_{-1.50}^{+1.56}$&$0.59_{-0.28}^{+0.27}$&$12.33_{-0.06}^{+0.07}$\\
$2.0<z<3.0$&\rule{0pt}{3ex}$-1.78_{-0.11}^{+0.06}$&$12.17_{-0.22}^{+0.19}$&$-1.31_{-0.33}^{+0.48}$&$2.81_{-1.43}^{+1.51}$&$0.59_{-0.35}^{+0.28}$&$12.50_{-0.08}^{+0.10}$
\enddata
\label{table:shmr_best_fit}
\tablecomments{$M_{\rm peak}$ is not a fitted parameter, but is derived from the others.}
\end{deluxetable*}
\capstarttrue
The SHMR is of great importance to galaxy formation models as it can be interpreted as the star formation history integrated over the lifetime of the halo, and thus as the efficiency with which baryons are converted into stars in halos of a given mass (assuming a constant fraction of a halo's mass is composed of baryons).

We are able to constrain this relation for our two lowest redshift bins ($1.5<z<2.0$ and $2.0<z<3.0$) as shown in \autoref{fig:shmr}, where we plot the ratio of the median stellar mass of a galaxy sample to the characteristic halo mass $M_{\rm h,min}$. The error bars shown here have been propagated from the uncertainty on $M_{\rm h,min}$ derived from our MCMC fitting procedure. At higher redshifts, our clustering measurements lack the range of stellar/halo masses required to adequately characterize this relationship.

At $4.0<z<5.0$ our data suggest an increase in the normalization of the SHMR. While this is a tentative result, we interpret this as subsequent stellar mass assembly being regulated by feedback processes, whilst the growth of a dark matter halo is not, reducing the normalization of this ratio over cosmic time. It also points toward the star formation efficiency being greater at higher redshifts.  

We describe the SMHR using the following functional form, proposed by Behroozi et al. (\citeyear{Behroozi:2013b}; see their Equation (4)):
\begin{equation}
\log_{10}(M_{\star}(M_{\mathrm{h}})) = \log_{10}(\varepsilon M_{\rm t}) + f\left(\log_{10}\left(\frac{M_{\mathrm{h}}}{M_{\rm t}}\right)\right)-f(0)\rm,
\label{eq:shmr}
\end{equation} 
where the function $f(x)$ is described by
\begin{equation}
f(x) = -\log_{10}(10^{\alpha x} + 1) + \delta\frac{(\log_{10}(1 + \exp(x)))^{\gamma}}{1 + \exp(10^{-x})}\rm.
\end{equation}

That is, a power law with slope $-\alpha$ for $M_{\rm h}\ll M_{\rm t}$ and an exponential transitioning into a sub-power law with slope $\gamma$ for $M_{\rm h}>M_{\rm t}$. Other parameterizations of this relationship have been proposed, mainly four-parameter double power-laws with different high- and low-mass slopes \citep[e.g.,][]{Yang:2003,Moster:2010}, however Behroozi et al. found that the five-parameter form performed better based on their abundance-matching technique. We fit this relation to our data, with the best-fit parameters being given in \autoref{table:shmr_best_fit}; the $1\sigma$ uncertainty of the fit is shown by the gray regions in \autoref{fig:shmr}. In Appendix~\ref{sec:smf} we show that we can accurately reconstruct the observed SMUVS stellar mass function using these best-fit relations, highlighting the consistency of our analysis. At $z\gtrsim3.0$ our results would be complemented by combining them with a consistent analysis of data from larger and/or deeper surveys in order to fully resolve the SHMR.

The peaked nature of the SHMR is commonly interpreted as the integrated effect of stellar and AGN feedback processes inhibiting star formation in the low- and high-mass regimes respectively within the hierarchical structure formation of $\Lambda$CDM \citep[e.g.,][]{Benson:2003,Bower:2006}. Given its peaked nature, the halo mass at which the SHMR is maximized, $M_{\rm peak}$, is of further interest. This indicates the halo mass at which the conversion of baryons into stars over the history of the halo has been most efficient. We derive this from the maximum of our best-fit SHMRs and give our values for $M_{\rm peak}$ in \autoref{fig:mpeak_z} (see also the rightmost column of \autoref{table:shmr_best_fit}). We can see that our estimations link up smoothly with previous estimates from similar studies at lower redshifts. Though we note that the \cite{Zehavi:2011} and \cite{Coupon:2012} studies are based on luminosity-selected samples, rather than stellar mass-selected.     
 
To describe the mild redshift dependence of $M_{\rm peak}$ we fit a simple power law to the observational data shown in \autoref{fig:mpeak_z} \citep[e.g.][]{Moster:2010}, and find that ${\log_{10}(M_{\rm peak}/\mathrm{M}_{\odot})=11.6\times(1 + z)^{0.06}}$ (shown as the solid line), is a reasonable description of the data, though this is of course somewhat dependent on the observational data chosen.

Our results for $M_{\rm peak}$ are consistent with those from the sophisticated abundance matching technique of \cite{Behroozi:2013b}, also shown in \autoref{fig:mpeak_z}. This model constrains the SHMR based on halo merger trees from the \emph{Bolshoi} simulations \citep{Klypin:2011} and observational estimates of the stellar mass function, the cosmic star formation rate density and the specific star formation rate -- stellar mass plane for $0<z<8$. Here we compute their value for $M_{\rm peak}$ based on their best-fit SHMR (see their Section~5) and propagate through their uncertainties on the SHMR parameters. Their model agrees well with other observational estimates of $M_{\rm peak}$ at low redshift ($z\lesssim1$), and is consistent with our measurements at higher redshifts, though understandably their model is less well constrained here due to the availability of observational data at these redshifts that could be used in their fitting procedure. 

\section{Discussion of Modeling Assumptions}
\label{sec:discussion}
In this section, we briefly discuss some issues pertaining to the validity of some of the assumptions we have made in our analyses. 

Throughout, we have assumed a five-parameter functional form for the HOD that is motivated by results from semi-analytical and hydrodynamical simulations \citep[e.g.,][]{Zheng:2005,Zheng:2007}. As is discussed in \cite{Guo:2016}, this follows from assuming a lognormal distribution for central galaxy stellar mass at a fixed halo mass, and a power-law relation between the mean stellar mass of central galaxies and their host halo masses. In regimes where the SHMR deviates significantly from a power law, this halo occupation may not hold. \cite{Leauthaud:2011} investigated this and found that changing the form of their assumed SHMR affected their best-fit HOD parameters only within their $1\sigma$ uncertainty. Thus while introducing different functional forms for the HOD may change the interpretation of some of the parameters the modeling results should not be significantly affected. We note that other forms for the galaxy HOD have been proposed in the literature \citep[e.g.,][]{Geach:2012,GonzalezPerez:2017}, though these are often designed to model populations that have been selected by properties that trace star formation, rather than stellar mass.  

We include an additional free parameter, $\Delta_{z}$, in order to encapsulate the dispersion due to the use of photometric redshifts in our analysis. Here we have assumed that this results in a Gaussian dispersion that evolves linearly with $(1+z)$, leading us to assume the window function in \autoref{eq:sigma_z_window} when projecting our spatial correlation functions according to \autoref{eq:limber}. These assumptions are based on comparisons of our `best-fit' {\sc LePhare} photometric redshifts, with literature spectroscopic redshifts in COSMOS. The distribution of the resulting $(z_{\rm phot}-z_{\rm spec})/(1 + z_{\rm spec})$ roughly resembles a Gaussian. However, there is a small tail of outliers that we do not account for explicitly in our analysis. \cite{Benjamin:2010} present a method that can in principle be used to estimate the impact that outlying photometric redshifts can have on clustering analyses. However, this method does not provide a unique solution for redshift bin contamination due to photometric redshift outliers because of degeneracies in the parameter space. We therefore refrain from implementing the method of Benjamin et al. and note that we do not place any prior constraint on the value of $\Delta_{z}$ determined by our fitting procedure. In addition, these outliers are a very small proportion ($\sim5$~percent) of the sample for which we have reliable spectroscopic redshifts, so we expect the impact of these on our science results to be minimal and certainly sub-dominant to the more general redshift dispersion which we have accounted for. In Appendix~\ref{sec:delta_z_discuss} we show that we can reasonably reproduce the observed photometric redshift distribution, i.e., one constructed from the best-fit photometric redshifts computed by {\sc LePhare}, from our best-fit values of $\Delta_{z}$, indicating the consistency of this method, and that values of $\Delta_{z}(1+z_{50})/(z_{\rm hi}-z_{\rm lo})$ are typically $\lesssim0.5$.

In implementing our HOD model we make a number of further assumptions. One of which is that the distribution of satellite galaxies traces that of the dark matter profile within a halo, which is assumed to be of the \cite{NFW:1997} form. This assumption may affect our conclusions regarding the satellite fraction of galaxies. Again, it is motivated to some degree by hydrodynamical simulations \citep[e.g.,][]{NagaiKravtsov:2005} though some studies have shown that this can be sensitive to the implementation of baryonic physics \citep[e.g.,][]{Simha:2012}.

The halo bias parameterization we are using \citep{Tinker:2010} is calibrated against $N$-body simulations for $z\lesssim2.5$, so there is some uncertainty about the application to the higher redshifts studied here. Additionally, more detailed modeling of scale-dependent bias \citep[e.g.,][]{Angulo:2008} may be required at higher redshifts.  For example, nonlinear clustering \citep[e.g.,][]{Jose:2016} can provide a better fit to galaxy clustering at high redshifts ($z=3-5$, Jose et al.  \citeyear{Jose:2017}).               

Finally, the halo occupation approach used here implicitly assumes that galaxy occupation statistics depend solely on the mass of the host dark matter halos. Whilst it is generally accepted that this is the halo property that most significantly influences the properties of the galaxies they host, over the last decade or so a body of evidence has emerged suggesting that it may also depend on additional properties such as the halo formation time \citep[e.g.,][]{Gao:2005, Croton:2007, Zenter:2014}.  This phenomenon is often referred to as `halo assembly bias', and adjustments to the standard HOD formalism in order to account for this have been proposed \citep[e.g.,][]{Hearin:2016}. However, it is still not clear if assembly bias is a significant effect, either in hydrodynamical simulations \citep[e.g.,][]{Chaves-Montero:2016}, or observations \citep[e.g.,][]{Lin:2016}, so we do not consider it here.  
\section{Conclusions}
\label{sec:conclusion}
In this work we have applied a phenomenological halo model to the observed clustering and abundance of galaxies in the \emph{Spitzer} Matching survey of the UltraVISTA ultra-deep Stripes (SMUVS) to understand the connection between galaxies and their host dark matter halos. SMUVS provides a unique combination of large area ($\sim0.66$ square degrees of the COSMOS field) and unparalleled depth at $3.6$ and $4.5$~$\mu$m. This, combined with the latest ultra-deep UltraVISTA data and other ancillary data in the COSMOS field, allows for precise photometric estimates of redshift and stellar mass and robust statistics from such a large catalog ($\sim2.9\times10^5$ objects). As a result, our analysis is performed over an unprecedented redshift range ($1.5<z<5.0$) for volume-limited galaxy samples selected by stellar mass. 

To interpret the observed clustering we utilize a well-established five-parameter halo model, though we fix the values of three of these parameters. Novel to our approach is the inclusion of the photometric redshift dispersion, $\Delta_{z}$, as a free parameter in the fitting procedure. This parameter is designed to encapsulate the effect of dispersion in the photometric redshifts used in our analysis and allows us to fully exploit the clustering information in our photometric galaxy catalog. This parameter is discussed in more detail in Appendix~\ref{sec:delta_z_discuss}. 

From our best-fit halo models we can derive a number of interesting quantities. The characteristic halo masses, $M_{\rm h,min}$ and $M_{\rm h,1}$, are well constrained by our data and exhibit fairly tight linear relationships (in logarithmic space) with stellar mass. This is similar to what has been found in earlier studies with similar analyses \citep[e.g.,][]{McCracken:2015}.

The satellite fraction, i.e., the fraction of galaxies in a sample that do not sit at the center of the potential well of their halo, generally increases with cosmic time. This is in broad agreement with other studies in the literature and in line with expectations of hierarchical structure formation.

The comoving correlation length, $r_{0}$, [defined such that $\xi(r_{0})\equiv1$], and large-scale effective galaxy bias, $b_{\rm gal}$, show strong monotonic trends with increasing redshift for a fixed stellar mass over the entire redshift range probed.

Projecting $\Lambda$CDM halo mass accretion histories computed according to Appendix A of \cite{McBride:2009} through our measurements of the large-scale bias, and combining this with the evolution of stellar mass content based on the cumulative number density \citep{Behroozi:2013a}, we propose an evolutionary path for the dark matter and stellar mass content of galaxies for $1.5<z<4.0$. This shows a favorable agreement with our independent determinations of the relationship between galaxy bias and stellar mass at these redshifts and appears to be an interesting method to investigate the coevolution of these quantities in the future.

Finally, we investigate the SHMR. This indicates the efficiency with which baryons have been converted into stars over the lifetime of a dark matter halo. We fit the SHMR parametrization of \cite{Behroozi:2013b} for our two lowest redshift bins ($1.5<z<2.0$ and $2.0<z<3.0$). In Appendix~\ref{sec:smf} we show that these SHMRs can reproduce the stellar mass function computed directly from the SMUVS catalog, highlighting the consistency of our analysis. The halo mass at which this relationship peaks indicates the halos for which the integrated star formation efficiency is greatest. We conclude that this quantity evolves mildly with redshift, i.e., $\log_{10}(M_{\rm peak})\propto(1+z)^{0.06}$ for $z\lesssim3$. Our inferred values of $M_{\rm peak}$ are consistent with independent constraints on the evolution of this quantity inferred from abundance-matching techniques \citep{Behroozi:2013b}.                   
  
We end by acknowledging that the results presented in this work could be complemented in the future by deeper observations and/or larger area surveys. This would allow a number of trends/relationships identified here (and in earlier studies) to be fully investigated at higher redshift, which will provide important constraints for physical models of galaxy formation and evolution.  Conversely, further modeling of our data with more physically motivated models than the HOD framework used in this study would allow us to disentangle the physical processes responsible for the trends observed. 
\section*{Acknowledgements}
The authors would like to thank the anonymous referee for a comprehensive and constructive report that allowed us to improve the overall quality of the manuscript. This work is based in part on observations carried out with the \emph{Spitzer Space Telescope}, which is  operated by the Jet Propulsion Laboratory, California Institute of Technology under a contract with NASA. It also uses data products from observations conducted with ESO Telescopes at the Paranal Observatory under ESO program ID 179.A-2005, data products produced by TERAPIX and the Cambridge  Astronomy Survey Unit on behalf of the UltraVISTA consortium, observations carried out by NASA/ESA \emph{Hubble Space Telescope} obtained and archived at the Space Telescope Science Institute; and the Subaru  Telescope, which is operated by the National Astronomical Observatory of Japan. This research has made use of the NASA/IPAC Infrared Science Archive, which is operated by the Jet Propulsion Laboratory, California Institute of Technology, under contract with NASA. The authors would like to thank Cedric Lacey and the \texttt{GALFORM} collaboration for use of the \texttt{GALFORM} model in Appendix~\ref{sec:delta_z_discuss}. This work has made use of the open-source \textsc{Python} packages: \texttt{Numpy}, \texttt{Scipy}, \texttt{Matplotlib} and \texttt{IPython}. The color-vision-impaired-friendly color schemes used throughout can be found at \url{https://personal.sron.nl/~pault/}. 

W.I.C., S.D., and K.I.C. acknowledge funding from the European Research Council through the award of the Consolidator Grant ID 681627-BUILDUP.
\appendix
\section{Modeling the photometric redshift dispersion}
\label{sec:delta_z_discuss}
\begin{figure*}
\centering
\includegraphics[trim = 0 0 0 0,clip = True, width = 0.99\linewidth]{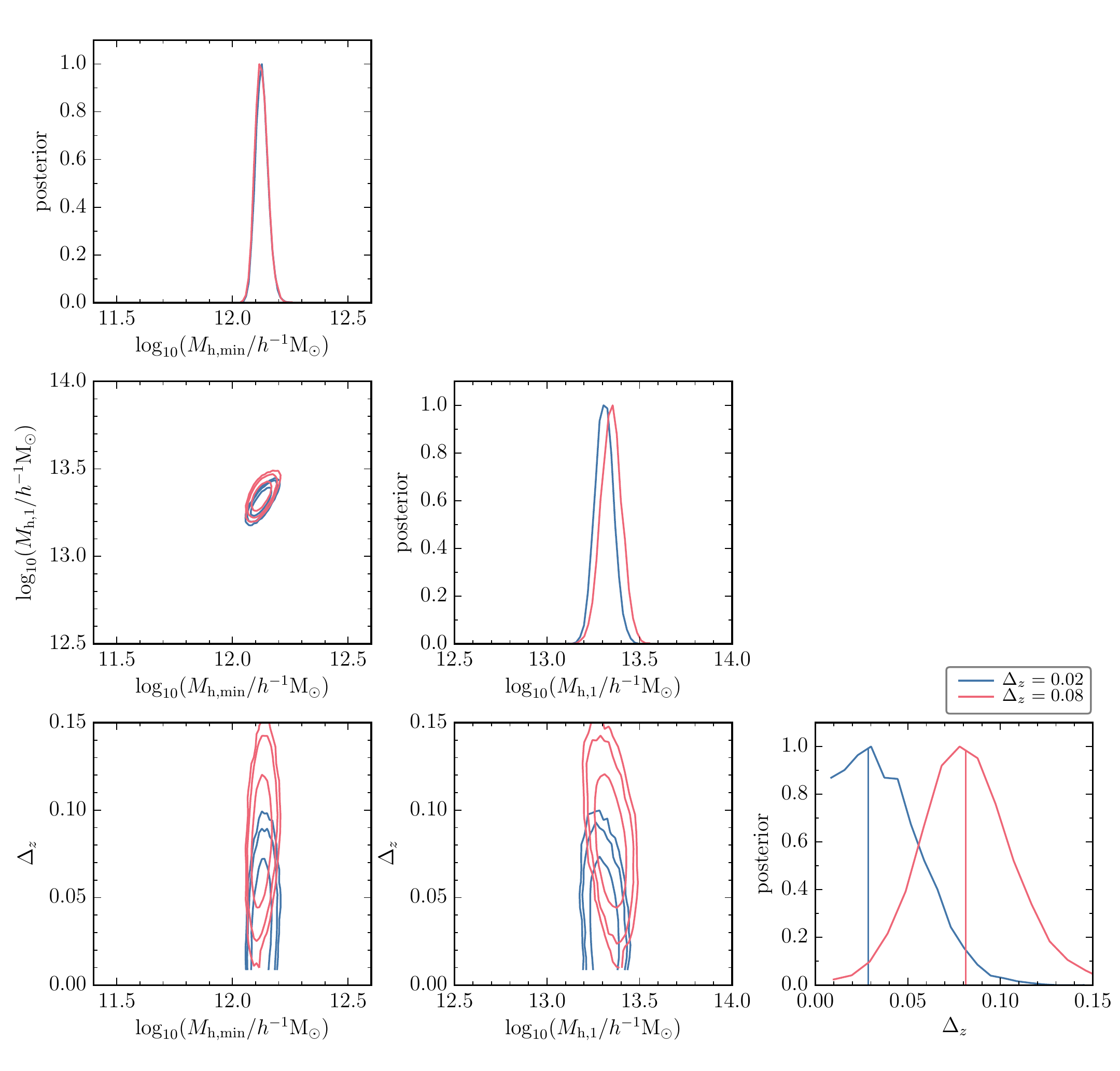}
\caption{Example one-dimensional (diagonal panels) and two-dimensional (off-diagonal panels) likelihood distributions produced from our MCMC fitting procedure for $\log_{10}(M_{\star}/\mathrm{M}_{\odot})>10.4$ galaxies in the range $2.0<z_{\rm phot}<3.0$, based on the clustering of a $2$~deg$^{2}$ simulated galaxy catalog produced by the \cite{Lacey:2016} \texttt{GALFORM} model. The different colors represent different values of input photometric redshift dispersion, $\Delta_{z}$, as indicated in the legend.  The solid lines in the bottom right panel indicate the best-fit value of $\Delta_{z}$, which is in remarkable agreement with the input value. The contours in the off-diagonal panels indicate the $1$, $2$ and $3$~$\sigma$ regions.}
\label{fig:corner_galform_delta_z}
\end{figure*}
\begin{figure*}
\centering
\includegraphics[width = 8.6189cm]{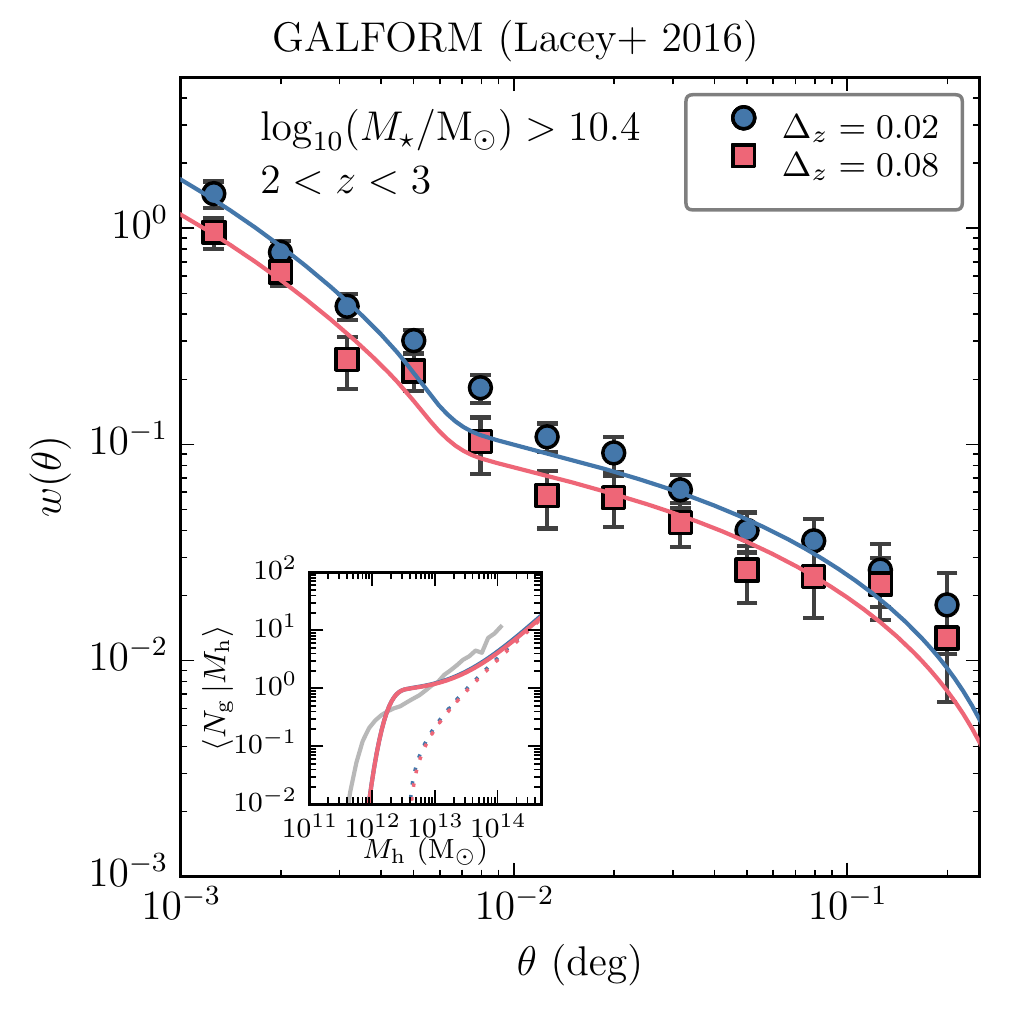}\hspace{0.79173cm}\includegraphics[width = 8.6189cm]{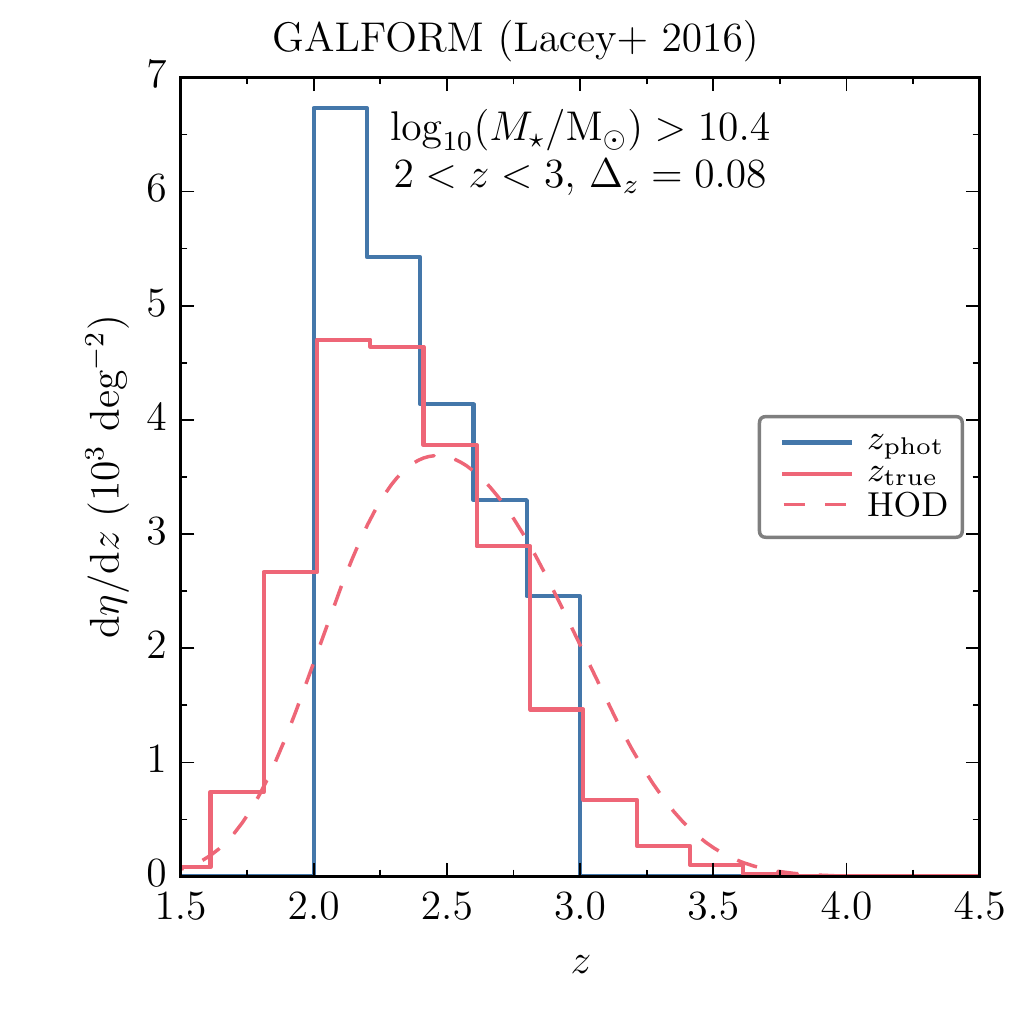}
\caption{\emph{Left panel}: the angular correlation function measured from a $2$~deg$^{2}$ simulated galaxy catalog produced by the \cite{Lacey:2016} \texttt{GALFORM} model (points with error bars) for simulated galaxies with $2.0<z_{\rm phot}<3.0$ and $\log_{10}(M_{\star}/\mathrm{M}_{\odot})>10.4$. The solid lines indicate the best-fit halo model, according to our modeling procedure outlined in Section~\ref{sec:methods}. The resulting HODs are shown in the inset panel; the HOD predicted by the \texttt{GALFORM} model is also shown here for reference (gray line).  The different colors represent different values of input photometric redshift dispersion, as indicated in the legend. \emph{Right panel}: shows the photometric and true redshift distributions (blue and red histograms respectively) for galaxies selected with $2.0<z_{\rm phot}<3.0$ and $\log_{10}(M_{\star}/\mathrm{M}_{\odot})>10.4$ with $\Delta_{z}=0.08$. The best-fit HOD reconstruction of the true redshift distribution is shown as the dashed red line.}
\label{fig:wtheta_hod_dndz_galform}
\end{figure*}
\begin{figure}
\centering
\includegraphics[width = 8.6189cm]{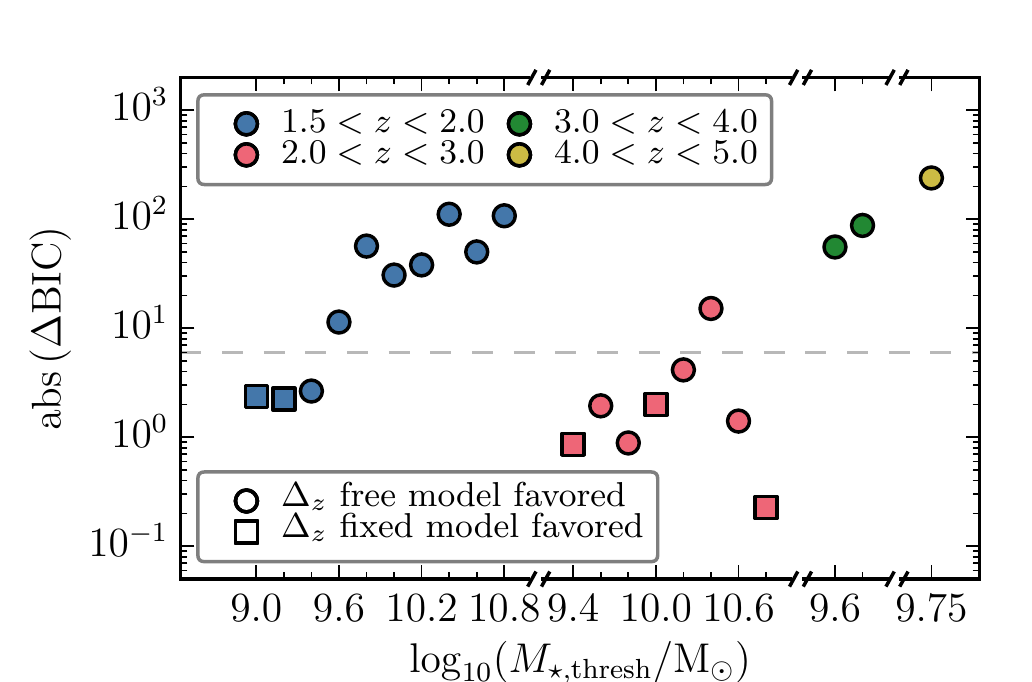}\hspace{0.79173cm}\includegraphics[width = 8.6189cm]{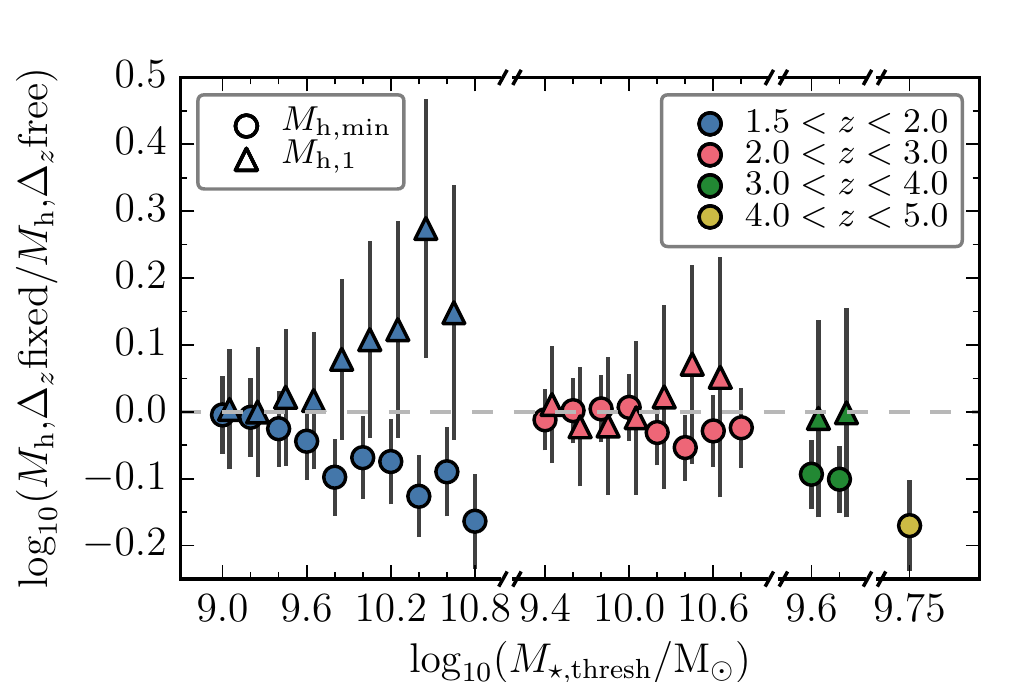}
\caption{\emph{Left panel}: difference in the Bayesian information criterion (BIC) between models in which $\Delta_{z}$ is fixed to a value of $0.035$, as suggested by comparing our spectroscopic and photometric redshifts, to models where it is a free parameter. A circle indicates that the $\Delta_{z}$ free model is favored, i.e., has the minimum BIC, whereas a square indicates that the $\Delta_{z}$ fixed model is favored. The different colors represent different photometric redshift bins as indicated in the legend. For clarity, samples are displayed from left to right in order of increasing stellar mass threshold within each redshift bin. For reference, a dashed line is drawn at a value of $\Delta\mathrm{BIC}=6$. A value of $\Delta\mathrm{BIC}>6$ indicates that this model is `strongly' favored. The \emph{right panel} shows the fractional difference on the derived best-fit characteristic halo masses, $M_{\rm h,min}$ (circles) and $M_{\mathrm{h,}1}$ (triangles), resulting from fixing $\Delta_{z}=0.035$, compared to leaving it as a free parameter in our model fitting. For clarity, the results for each sample are displayed as described for the left panel.}
\label{fig:BIC_dmhs}
\end{figure}
\begin{figure}
\centering
\includegraphics[width = 8.6189cm]{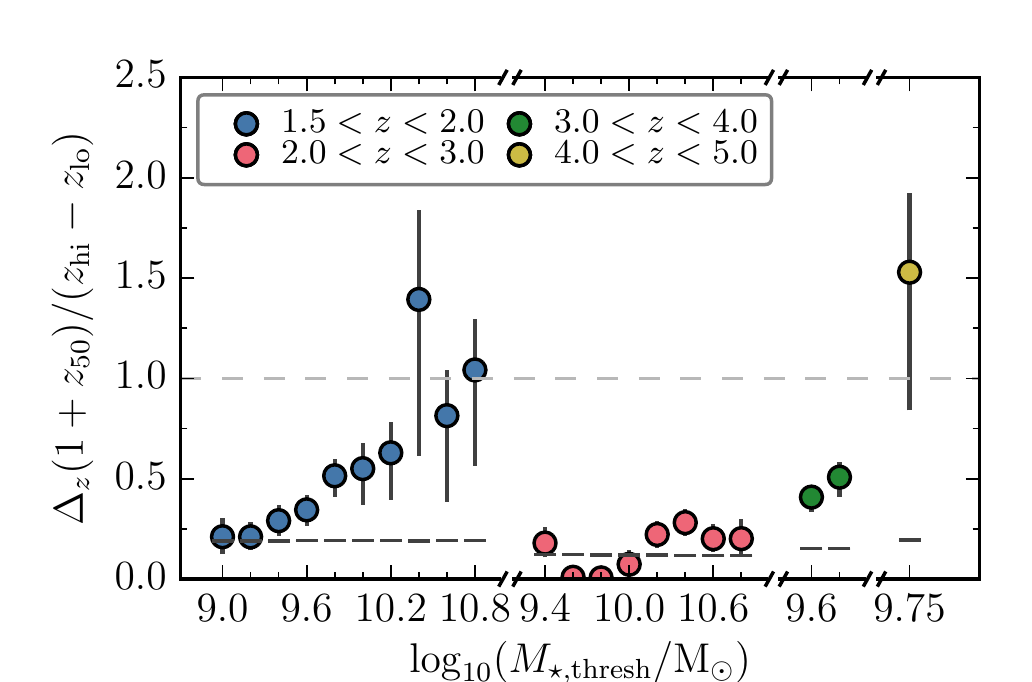}
\caption{Best-fit absolute redshift dispersion as a fraction of the photometric redshift bin width (here $z_{50}$ is the median stellar mass of a sample and $z_{\rm hi}$ and $z_{\rm lo}$ represent the limits of the photometric redshift bin). For clarity, the results for each of our samples are displayed as described for \autoref{fig:BIC_dmhs}. The black dashed lines indicate a value of $\Delta_{z}=0.035$, suggested by comparing our spectroscopic and photometric redshifts. For reference, a horizontal gray dashed line is drawn at unity.}
\label{fig:delta_z_abs}
\end{figure}
\begin{figure}
\centering
\includegraphics[trim = 0 25 0 25,clip = True, width = \linewidth]{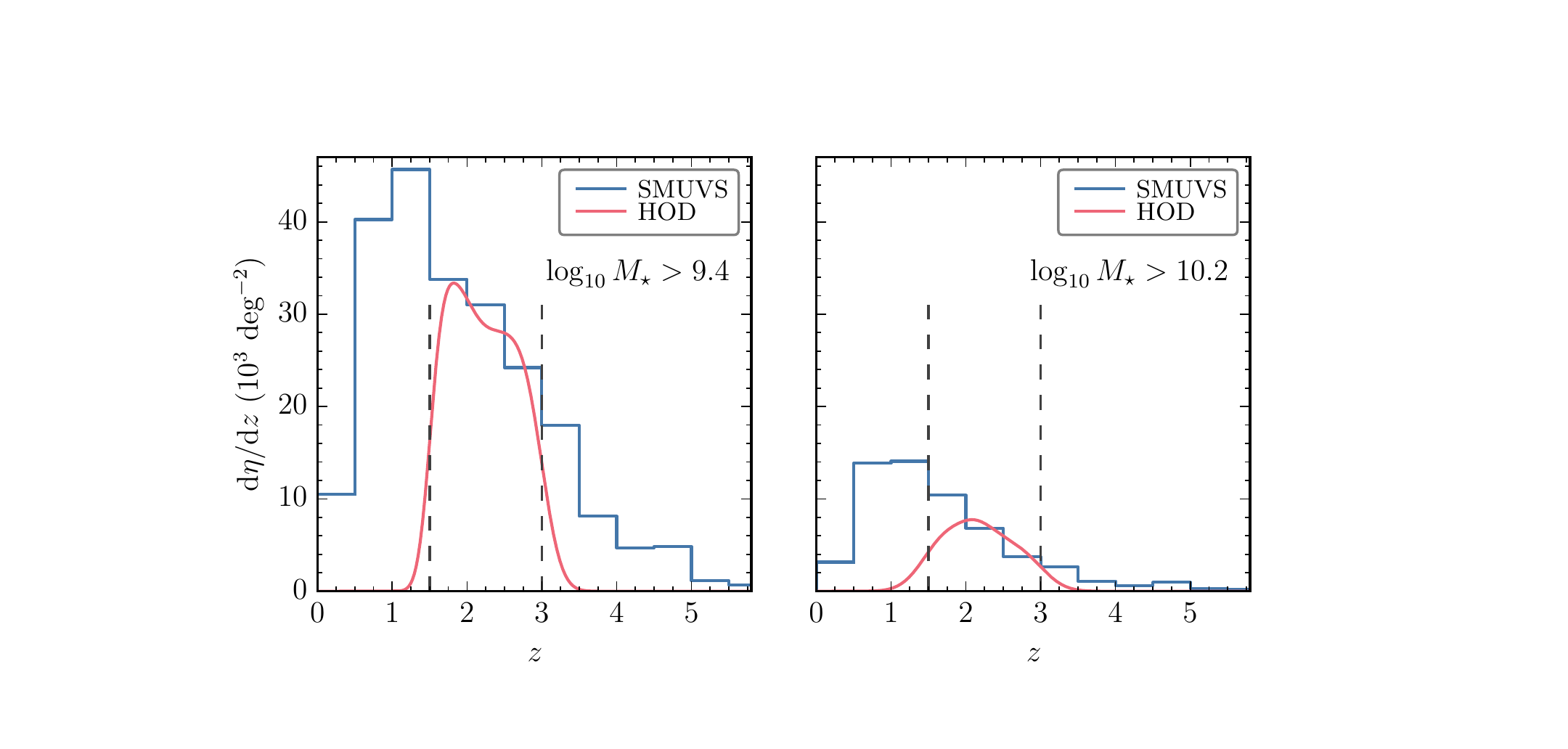}
\caption{Example photometric redshift distribution of SMUVS galaxies (blue lines) and the best-fit HOD reconstruction (red lines) in the range $1.5<z<3.0$ (indicated by the vertical dashed lines). Stellar mass thresholds used are $\log_{10}(M_{\star}/\mathrm{M}_{\odot})>9.4$ (left panel) and $\log_{10}(M_{\star}/\mathrm{M}_{\odot})>10.2$ (right panel).}
\label{fig:dndz_hod}
\end{figure}
In this appendix, we discuss our modeling of the redshift dispersion caused by using photometric redshifts, which we describe through the parameter $\Delta_{z}$ and \autoref{eq:sigma_z_window}. This assumes that the distribution of $(z_{\rm phot}-z_{\rm true})$ is Gaussian with a width that scales with $(1+z_{\rm true})$. First, we show that under these assumptions, our modeling can accurately recover the redshift dispersion. To do this we use a simulated $2$~deg$^{2}$ galaxy catalog from the semi-analytical model of galaxy formation, \texttt{GALFORM} \citep{Lacey:2016}. We simulate photometric redshifts by applying a Gaussian scatter that scales with $(1+z_{\rm true})$ and then select galaxies based on their photometric redshifts in the range $2<z_{\rm phot}<3$ and $\log_{10}(M_{\star}/\mathrm{M}_{\odot})>10.4$. We do this for two values of standard deviation for the Gaussian scatter, $\Delta_{z}$, of $0.02$ and $0.08$. For simplicity, we do not apply any scaling or error to the simulated stellar masses.  We then measure the angular clustering and fit a HOD model to this following the same procedure we use for our SMUVS data as outlined in Section~\ref{sec:methods}, for this we assume a $10$~percent error on $N_{\rm gal}$ as is typical of our SMUVS data. The resulting marginalized likelihood distributions from our MCMC fitting are shown in \autoref{fig:corner_galform_delta_z}.  We can see from the bottom right panel that the best-fit MCMC value for $\Delta_{z}$ is remarkably close to the input value, and the other panels show that the impact on the other two halo model parameters, $M_{\rm h,min}$ and $M_{\mathrm{h,}1}$, is minimal. 

The effect of increasing the photometric redshift dispersion on the measured clustering is shown in the left panel of \autoref{fig:wtheta_hod_dndz_galform}. We can see that increasing the value of $\Delta_{z}$ reduces the amplitude, as the clustering is being projected over a broader redshift range. However, the model is able to account for this such that the best-fit HOD that is recovered is the same, as can be seen in the inset panel. In the inset panel we also show the HOD predicted by the \texttt{GALFORM} model. We note that this is not precisely reproduced by the statistical HOD model, but consider further investigation of this beyond the scope of this work. The photometric redshift broadening is shown further in the right panel of \autoref{fig:wtheta_hod_dndz_galform} for the case of $\Delta_{z}=0.08$. We see here that the true redshift distribution of galaxies selected by $2<z_{\rm phot}<3$ is actually somewhat broader than these photometric redshift limits suggest and that this can be reasonably recovered by the HOD model, reminding the reader that the HOD model cannot account for evolution of the population within the redshift bin chosen.  We conclude from these tests that our methodology can accurately account for the redshift dispersion arising from our necessary use of photometric redshifts.

We now discuss our decision to leave $\Delta_{z}$ as a free parameter. In principle, $\Delta_{z}$ could be measured from the data by comparison with spectroscopic redshifts. However, the spectroscopic coverage of our catalog is small ($\sim5$~percent) and has a complicated selection function.  Nevertheless, our data suggest $\sigma_{z}=0.035$ for galaxies in the redshift range covered by this study $(1.5<z<5.0)$. We remind the reader that $\sigma_{z}$ is the standard deviation of $|z_{\rm phot}-z_{\rm spec}|/(1+z_{\rm spec})$, excluding outliers [objects with $|z_{\rm phot}-z_{\rm spec}|/(1+z_{\rm spec})>0.15$]. We therefore fit a set of HOD models fixing $\Delta_{z}=0.035$ and compare the resulting best-fit models to our fits with $\Delta_{z}$ left as a free parameter. To do this, we use the Bayesian information criterion (BIC; Schwarz \citeyear{Schwarz:1978}; see e.g., \citealt{Liddle:2007} for a discussion). This criterion is evaluated as follows:
\begin{equation}
\mathrm{BIC} = -2\ln\mathcal{L}_{\rm max} + k\ln N\rm,
\label{eq:BIC} 
\end{equation}  
where $k$ is the number of model parameters and $N$ is the number of data points ($12$ angular clustering data points and the observed number of galaxies). It is an approximation of the Bayesian evidence that should be a reasonable one given our uninformative priors. Models with the lowest value of BIC are favored, and though models with more free parameters can generally achieve a greater maximum likelihood, they are penalized by the $k\ln N$ term in \autoref{eq:BIC}. We consider the difference in BIC values, $\Delta\mathrm{BIC}$, between our $\Delta_{z}$ fixed and $\Delta_{z}$ free models in the left panel of \autoref{fig:BIC_dmhs}. As a `rule of thumb', values of $\Delta\mathrm{BIC}>6$ indicate that a model is `strongly' favored \citep[e.g., Appendix E of][]{Fabozzi:2014}. Thus, we can see that in many cases the models in which $\Delta_{z}$ is a free parameter are strongly favored.  

Additionally, we assess the impact of $\Delta_{z}$ on the two remaining halo model parameters. The fractional change in these is typically $\lesssim0.2$~dex, as is shown in the right panel of \autoref{fig:BIC_dmhs}. Interestingly, a higher value of $\Delta_{z}$ appears to produce slightly higher values of $M_{\rm h,min}$ (which will lead to higher values of derived quantities such as $b_{\rm gal}$ and $r_{0}$) and slightly lower values of $M_{\mathrm{h,}1}$ (which will lead to higher values of $f_{\rm sat}$). Given that $\Delta_{z}$-free models are strongly favored by the BIC and the impact on the other model parameters is fairly small ($\lesssim0.2$~dex) we are justified in our decision to leave $\Delta_{z}$ as a free parameter.

Finally, we discuss whether our modeling returns reasonable values for $\Delta_{z}$. In \autoref{fig:delta_z_abs} we show our values of $\Delta_{z}\,(1+z_{50})/(z_{\rm hi}-z_{\rm lo})$, which gives a measure of the redshift dispersion relative to the width of the redshift bin we are considering (here $z_{50}$ is the median stellar mass of a sample and $z_{\rm hi}$ and $z_{\rm lo}$ represent the limits of the photometric redshift bin). The values of this quantity are typically $\lesssim0.5$, which suggests that the bin-to-bin photometric redshift contamination in our work is minor once the general redshift dispersion has been accounted for. We note that our $1.5<z<2.0$, $\log_{10}(M_{\star}/\mathrm{M}_{\odot})>10.4$ sample appears to return an anomalously high value of $\Delta_{z}$, however, none of our scientific conclusions would be changed by removing this sample from our study. Our sample at $4.0<z<5.0$ also returns a high value for $\Delta_{z}$, which we accept as a consequence of the extreme high-redshift nature of this sample in terms of stellar mass-selected clustering measurements.

A further check on the values of $\Delta_{z}$ returned by our model is to use them to reconstruct the photometric redshift distribution of our catalog for a given stellar mass threshold. We show this in \autoref{fig:dndz_hod} over the redshift range $1.5<z<3.0$ for two stellar mass thresholds. These appear to be a reasonable reconstruction of the redshift distribution computed directly from the SMUVS catalog, which is based on the best-fit photometric redshift computed by {\sc LePhare}, and so add confidence to the values of $\Delta_{z}$ determined by our method.

\section{Reconstructing the stellar mass function}
\label{sec:smf}
\begin{figure*}
\centering
\includegraphics[trim = 0 25 0 25,clip = True, width = \linewidth]{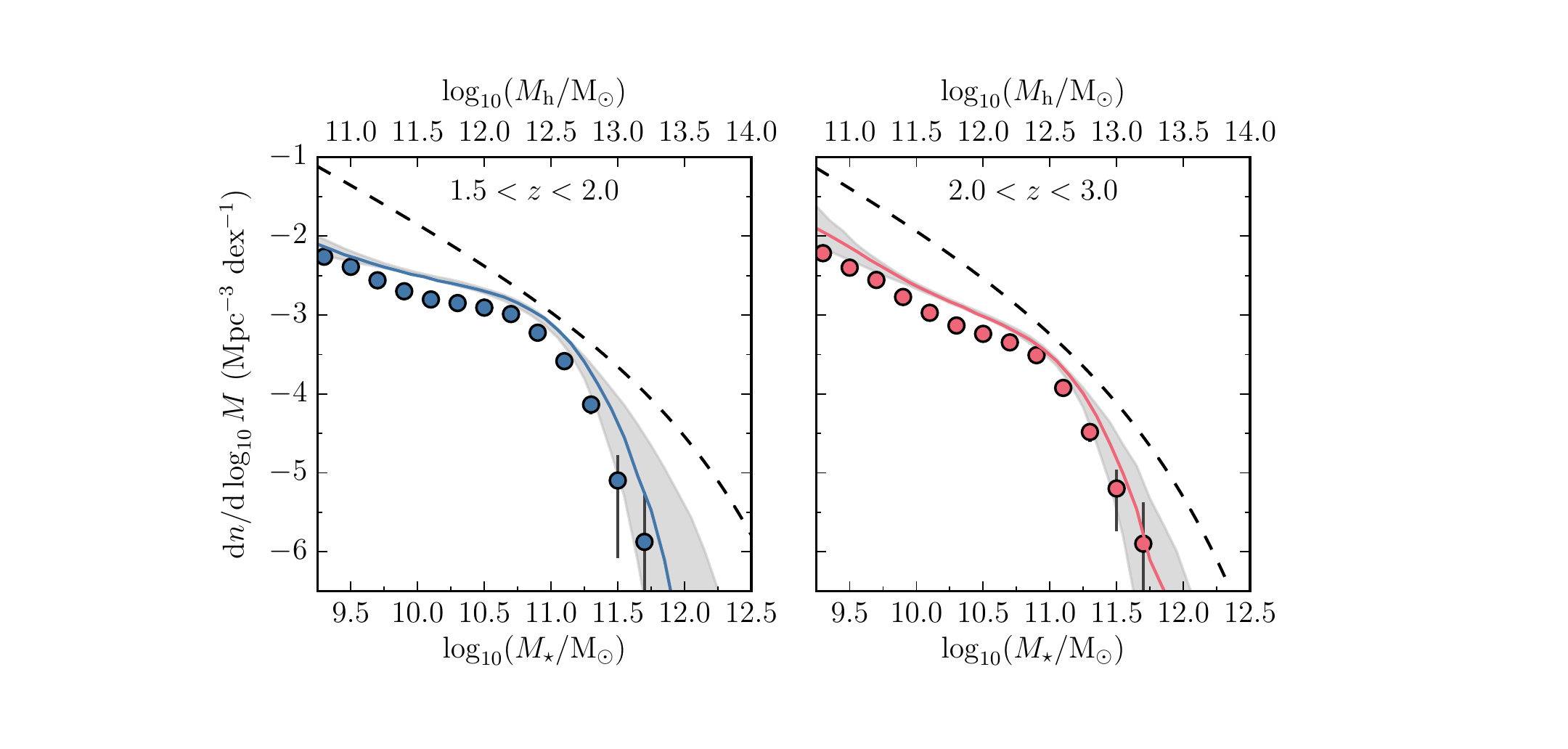}
\caption{The SMUVS stellar mass functions for $1.5<z<2.0$ (left panel) and $2.0<z<3.0$ (right panel).  The solid lines indicate this statistic computed using the best-fit SHMRs from Section~\ref{sec:shmr} (see \autoref{table:shmr_best_fit}), with the propagated $16-84$ percentile errors indicated by the gray shaded region. For reference, the dashed line indicates the halo mass function of Tinker et al. (\citeyear{Tinker:2008}). The points with error bars indicate the stellar mass function computed directly from the SMUVS catalog as described by \cite{Deshmukh:2018}.}
\label{fig:smf}
\end{figure*}
In this appendix, we use our best-fit SHMRs from Section~\ref{sec:shmr} (see \autoref{table:shmr_best_fit}) to reconstruct the stellar mass function for $1.5<z<2.0$ and $2.0<z<3.0$. For this, we begin with a halo mass function generated according to the prescription of \cite{Tinker:2008} and incorporate a constant Gaussian scatter around our best-fit SHMR with a dispersion of $0.16$~dex \citep{More:2009,Moster:2010}. We show the results in \autoref{fig:smf}, where we compare our SHMR-derived stellar mass function to those computed directly from the SMUVS catalog using the method described in \cite{Deshmukh:2018}. The two methods are in good agreement, highlighting the consistency of our analyses. One can see clearly how the peaked nature of the SHMR converts the halo mass function to the Schechter function shape of the observed galaxy stellar mass function. We have ignored satellite galaxies in converting a halo mass function to a stellar mass function as shown in \autoref{fig:smf}. However, as our satellite fractions are generally small ($\lesssim0.1$) this does not affect the conclusion of this section.
\section{Tabulated Results}
In this appendix, we tabulate our main results from Section~\ref{sec:results} in \autoref{table:main_results}.
\label{sec:main_results_table}
\begin{sidewaystable}
\centering
\caption{Tabulated main results}
\label{table:main_results}

\begin{tabular}{rrrrrrrrrrrrrrrrrrr}\toprule
\rule{0pt}{3ex}$z_{\rm lo}$ & $z_{\rm hi}$ & $M_{\star\mathrm{,lim}}$ & $z_{50}$ & $M_{\star\mathrm{,}50}$ & $N_{\rm gal}$ & $\sigma_{N_{\rm gal}}$ & $\sigma_{\sqrt{N}}$ & $\sigma_{\rm CV}$ & $\sigma_{\rm fit}$ & $M_{\rm h,min}$ & $M_{\rm h,1}$ & $\Delta_{z}$ & $f_{\rm sat}$ & $r_{0}$ & $b_{\rm gal}$ & $\chi^{2}_{w(\theta)}$ & $\chi^{2}_{N_{\rm gal}}$ & $\chi^{2^{\dagger}}_{\rm red}$ \\ \cmidrule{8-10}
 & & & & & & &\multicolumn{3}{c}{(percentage error)} & & & & & ($h^{-1}$~Mpc) & & & & \\
\midrule
\rule{0pt}{3ex}$1.5$&$2.0$&$ 9.00$&$1.70$&$ 9.49$&$19\,637$&$1\,424.3$&$0.71$&$7.21$&$0.40$&$11.58_{-0.04}^{+0.05}$&$12.88_{-0.07}^{+0.07}$&$0.039_{-0.017}^{+0.016}$&$0.06_{-0.01}^{+0.01}$&$4.14_{-0.10}^{+0.11}$&$1.87_{-0.03}^{+0.03}$&$42.74$&$0.012$&$4.75$\\ 
\rule{0pt}{3ex}&&$ 9.20$&$1.71$&$ 9.64$&$15\,309$&$1\,134.2$&$0.81$&$7.35$&$0.46$&$11.67_{-0.05}^{+0.05}$&$12.96_{-0.07}^{+0.08}$&$0.039_{-0.014}^{+0.013}$&$0.06_{-0.01}^{+0.01}$&$4.39_{-0.12}^{+0.13}$&$1.95_{-0.04}^{+0.04}$&$41.02$&$0.019$&$4.56$\\ 
\rule{0pt}{3ex}&&$ 9.40$&$1.72$&$ 9.86$&$11\,232$&$857.5$&$0.94$&$7.56$&$0.49$&$11.79_{-0.04}^{+0.05}$&$13.01_{-0.08}^{+0.08}$&$0.054_{-0.014}^{+0.014}$&$0.07_{-0.01}^{+0.01}$&$4.77_{-0.13}^{+0.14}$&$2.07_{-0.04}^{+0.04}$&$35.71$&$0.035$&$3.97$\\ 
\rule{0pt}{3ex}&&$ 9.60$&$1.73$&$10.09$&$ 8\,186$&$658.3$&$1.11$&$7.95$&$0.49$&$11.91_{-0.04}^{+0.05}$&$13.07_{-0.08}^{+0.09}$&$0.063_{-0.013}^{+0.015}$&$0.08_{-0.01}^{+0.02}$&$5.18_{-0.14}^{+0.17}$&$2.20_{-0.04}^{+0.05}$&$16.01$&$0.026$&$1.78$\\ 
\rule{0pt}{3ex}&&$ 9.80$&$1.73$&$10.27$&$ 6\,129$&$515.3$&$1.28$&$8.29$&$0.59$&$12.02_{-0.05}^{+0.05}$&$13.14_{-0.09}^{+0.10}$&$0.094_{-0.016}^{+0.019}$&$0.08_{-0.02}^{+0.02}$&$5.58_{-0.18}^{+0.19}$&$2.33_{-0.05}^{+0.06}$&$13.36$&$0.000$&$1.48$\\ 
\rule{0pt}{3ex}&&$10.00$&$1.74$&$10.42$&$ 4\,634$&$417.4$&$1.47$&$8.86$&$0.68$&$12.11_{-0.05}^{+0.06}$&$13.22_{-0.11}^{+0.12}$&$0.101_{-0.024}^{+0.033}$&$0.08_{-0.02}^{+0.03}$&$5.93_{-0.21}^{+0.23}$&$2.44_{-0.06}^{+0.07}$&$19.14$&$0.001$&$2.13$\\ 
\rule{0pt}{3ex}&&$10.20$&$1.73$&$10.54$&$ 3\,458$&$328.3$&$1.70$&$9.31$&$0.77$&$12.22_{-0.05}^{+0.06}$&$13.29_{-0.12}^{+0.13}$&$0.115_{-0.028}^{+0.043}$&$0.08_{-0.03}^{+0.04}$&$6.36_{-0.24}^{+0.26}$&$2.56_{-0.07}^{+0.08}$&$21.08$&$0.002$&$2.34$\\ 
\rule{0pt}{3ex}&&$10.40$&$1.72$&$10.66$&$ 2\,398$&$240.5$&$2.04$&$9.76$&$1.03$&$12.36_{-0.05}^{+0.06}$&$13.22_{-0.19}^{+0.15}$&$0.256_{-0.082}^{+0.143}$&$0.15_{-0.05}^{+0.10}$&$7.16_{-0.29}^{+0.39}$&$2.80_{-0.09}^{+0.12}$&$21.62$&$0.002$&$2.40$\\ 
\rule{0pt}{3ex}&&$10.60$&$1.74$&$10.79$&$ 1\,469$&$164.7$&$2.61$&$10.84$&$1.17$&$12.49_{-0.05}^{+0.06}$&$13.56_{-0.15}^{+0.15}$&$0.149_{-0.042}^{+0.078}$&$0.07_{-0.03}^{+0.04}$&$7.61_{-0.29}^{+0.33}$&$2.95_{-0.09}^{+0.10}$&$13.82$&$0.001$&$1.54$\\ 
\rule{0pt}{3ex}&&$10.80$&$1.75$&$10.95$&$  702$&$ 96.5$&$3.77$&$13.09$&$1.88$&$12.71_{-0.06}^{+0.07}$&$13.69_{-0.17}^{+0.16}$&$0.189_{-0.046}^{+0.087}$&$0.09_{-0.04}^{+0.06}$&$8.92_{-0.37}^{+0.45}$&$3.37_{-0.12}^{+0.14}$&$ 8.74$&$0.001$&$0.97$\\ \midrule
\rule{0pt}{3ex}$2.0$&$3.0$&$ 9.40$&$2.46$&$ 9.73$&$18\,362$&$12\,36.9$&$0.74$&$6.69$&$0.33$&$11.73_{-0.04}^{+0.04}$&$12.83_{-0.07}^{+0.07}$&$0.052_{-0.023}^{+0.020}$&$0.06_{-0.01}^{+0.01}$&$5.24_{-0.12}^{+0.13}$&$2.81_{-0.05}^{+0.05}$&$24.42$&$0.030$&$2.72$\\ 
\rule{0pt}{3ex}&&$ 9.60$&$2.46$&$ 9.92$&$12\,201$&$855.1$&$0.91$&$6.94$&$0.40$&$11.88_{-0.04}^{+0.04}$&$12.99_{-0.07}^{+0.07}$&$0.003_{-0.002}^{+0.027}$&$0.05_{-0.01}^{+0.01}$&$5.77_{-0.12}^{+0.13}$&$3.01_{-0.05}^{+0.05}$&$35.62$&$0.258$&$3.99$\\ 
\rule{0pt}{3ex}&&$ 9.80$&$2.45$&$10.14$&$ 7\,851$&$571.3$&$1.13$&$7.17$&$0.51$&$12.01_{-0.04}^{+0.04}$&$13.14_{-0.08}^{+0.08}$&$0.002_{-0.001}^{+0.032}$&$0.05_{-0.01}^{+0.01}$&$6.30_{-0.14}^{+0.16}$&$3.20_{-0.05}^{+0.06}$&$13.85$&$0.159$&$1.56$\\ 
\rule{0pt}{3ex}&&$10.00$&$2.42$&$10.35$&$ 5\,194$&$404.1$&$1.39$&$7.63$&$0.61$&$12.14_{-0.04}^{+0.04}$&$13.31_{-0.08}^{+0.09}$&$0.022_{-0.021}^{+0.029}$&$0.04_{-0.01}^{+0.01}$&$6.79_{-0.17}^{+0.20}$&$3.35_{-0.06}^{+0.07}$&$ 7.40$&$0.003$&$0.82$\\ 
\rule{0pt}{3ex}&&$10.20$&$2.40$&$10.52$&$ 3\,516$&$299.9$&$1.69$&$8.33$&$0.77$&$12.25_{-0.04}^{+0.04}$&$13.40_{-0.09}^{+0.11}$&$0.065_{-0.019}^{+0.020}$&$0.04_{-0.01}^{+0.01}$&$7.35_{-0.21}^{+0.24}$&$3.54_{-0.08}^{+0.09}$&$ 6.03$&$0.000$&$0.67$\\ 
\rule{0pt}{3ex}&&$10.40$&$2.38$&$10.66$&$ 2\,357$&$216.9$&$2.06$&$8.92$&$0.97$&$12.37_{-0.04}^{+0.04}$&$13.47_{-0.10}^{+0.11}$&$0.083_{-0.019}^{+0.019}$&$0.04_{-0.02}^{+0.02}$&$7.99_{-0.23}^{+0.26}$&$3.76_{-0.09}^{+0.10}$&$ 2.80$&$0.000$&$0.31$\\ 
\rule{0pt}{3ex}&&$10.60$&$2.35$&$10.81$&$ 1\,446$&$150.4$&$2.63$&$10.00$&$1.12$&$12.51_{-0.04}^{+0.05}$&$13.67_{-0.12}^{+0.15}$&$0.060_{-0.022}^{+0.020}$&$0.03_{-0.01}^{+0.02}$&$8.71_{-0.26}^{+0.32}$&$3.99_{-0.10}^{+0.12}$&$ 4.80$&$0.005$&$0.53$\\ 
\rule{0pt}{3ex}&&$10.80$&$2.33$&$10.93$&$  740$&$ 90.5$&$3.68$&$11.51$&$1.92$&$12.68_{-0.05}^{+0.06}$&$>13.81$&$0.061_{-0.030}^{+0.026}$&$<0.03$&$9.73_{-0.38}^{+0.41}$&$4.36_{-0.14}^{+0.16}$&$^{^{\dagger\dagger}}-$&$-$&$-$\\ \midrule
\rule{0pt}{3ex}$3.0$&$4.0$&$ 9.60$&$3.35$&$ 9.89$&$ 6\,187$&$612.5$&$1.27$&$9.80$&$0.57$&$11.81_{-0.04}^{+0.05}$&$12.97_{-0.10}^{+0.12}$&$0.094_{-0.014}^{+0.017}$&$0.02_{-0.01}^{+0.01}$&$6.52_{-0.20}^{+0.23}$&$4.13_{-0.09}^{+0.11}$&$14.46$&$0.011$&$1.61$\\ 
\rule{0pt}{3ex}&&$ 9.80$&$3.36$&$10.04$&$ 3\,894$&$398.3$&$1.60$&$10.08$&$0.69$&$11.93_{-0.04}^{+0.05}$&$13.07_{-0.10}^{+0.13}$&$0.116_{-0.017}^{+0.022}$&$0.02_{-0.01}^{+0.01}$&$7.10_{-0.21}^{+0.25}$&$4.41_{-0.10}^{+0.12}$&$ 6.54$&$0.000$&$0.73$\\ \midrule
\rule{0pt}{3ex}$4.0$&$5.0$&$ 9.75$&$4.56$&$10.02$&$ 1\,768$&$302.0$&$2.38$&$16.87$&$1.27$&$11.73_{-0.06}^{+0.07}$&$12.61_{-0.18}^{+0.18}$&$0.275_{-0.071}^{+0.124}$&$0.04_{-0.02}^{+0.04}$&$7.87_{-0.28}^{+0.39}$&$6.09_{-0.17}^{+0.23}$&$10.32$&$0.001$&$1.15$\\
\bottomrule
\multicolumn{19}{l}{\rule{0pt}{3ex}\textsc{Note.--} All masses in $\log_{10}(M/\mathrm{M}_{\odot})$. Error bars are $1\sigma$.}\\
\multicolumn{19}{l}{$^{\dagger}$Each of our fits has nine degrees of freedom.  There are $12$ clustering data points, plus the observed number of galaxies. The model has three free parameters.}\\\multicolumn{19}{l}{$^{\dagger\dagger}$For this sample, we are unable to fully constrain all of the model parameters, therefore we do not quote a minimum $\chi^{2}_{\rm red}$.}
\end{tabular}
\end{sidewaystable}
\section{Example Likelihood Distributions}
\label{sec:likelihood_example}
\begin{figure*}
\centering
\includegraphics[trim = 0 0 0 0,clip = True, width = 0.99\linewidth]{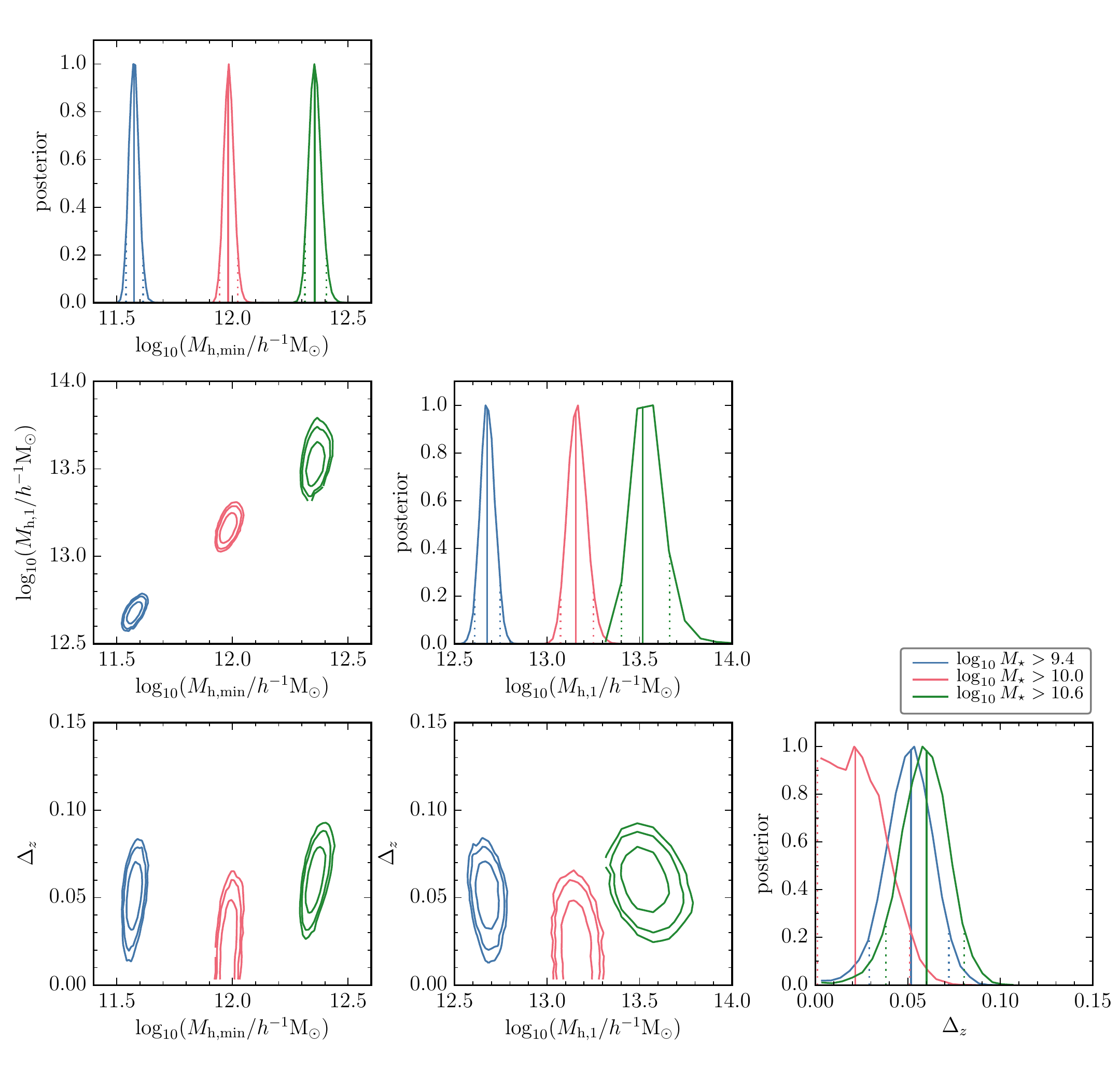}
\caption{Example one-dimensional (diagonal panels) and two-dimensional (off-diagonal panels) likelihood distributions produced from our MCMC fitting procedure for stellar mass-selected samples in the $2.0<z<3.0$ redshift range.  The contours in the off-diagonal panels represent the $1$, $2$, and $3$~$\sigma$ regions.  The solid and dashed lines in the diagonal panels represent the best-fit values and $1\sigma$ errors, respectively. The different colors correspond to a different stellar mass threshold as shown in the legend. The samples correspond to the clustering and best-fit halo models shown in the upper right-hand panel of \autoref{fig:wtheta_hod}.}
\label{fig:corner_example}
\end{figure*}

In this appendix, we show some examples of the likelihood distributions produced by our MCMC fitting procedure in \autoref{fig:corner_example}. These correspond to the samples shown in the upper right-hand panel of \autoref{fig:wtheta_hod}.

\bibliographystyle{apj}
\bibliography{ref}
\clearpage
\end{document}